\documentclass[11pt,letterpaper]{article}
 \pdfoutput=1

\usepackage{placeins}
\usepackage{geometry}
\usepackage[pdftex]{graphicx}
\usepackage{amssymb,amsfonts,amsmath}
\usepackage{color}
\usepackage{natbib}
\usepackage{hyperref}
\usepackage{amsthm,mathrsfs}
\usepackage[normalem]{ulem}
\usepackage{longtable}

\usepackage{xr}
\newcommand{\trans}{^{\mbox{\scriptsize {\sf T}}}}
\newcommand{\g}{\mathbf{g}}

\newcommand{\x}{\mathbf{x}}
\newcommand{\X}{\mathbf{X}}
\newcommand{\by}{\mathbf{y}}

\newcommand{\W}{\mathbf{W}}

\newcommand{\Z}{\mathbf{Z}}

\newcommand{\y}{\mathbf{y}}

\newcommand{\ba}{\mathbf{a}}
\newcommand{\bb}{\mathbf{b}}

\usepackage{color}
\usepackage{url}
\begin{document}
\title{{\sc IsoDOT} Detects Differential RNA-isoform Expression/Usage with respect to a Categorical or Continuous Covariate with High Sensitivity and Specificity}

\author{
Wei Sun, Yufeng Liu, 
James J. Crowley, 
Ting-Huei Chen, 
Hua Zhou,
Haitao Chu, \\
Shunping Huang,
Pei-Fen Kuan,
Yuan Li,
Darla Miller,
Ginger Shaw,
Yichao Wu,\\
Vasyl Zhabotynsky,
Leonard McMillan,
Fei Zou,
Patrick F. Sullivan,\\
and 
Fernando Pardo-Manuel de Villena
}

\maketitle

\begin{abstract} 
We have developed a statistical method named IsoDOT to assess differential isoform expression (DIE) and differential isoform usage (DIU) using RNA-seq data. Here isoform usage refers to relative isoform expression given the total expression of the corresponding gene. IsoDOT performs two tasks that cannot be accomplished by existing methods: to test DIE/DIU with respect to a continuous covariate, and to test DIE/DIU for one case versus one control. The latter task is not an uncommon situation in practice, e.g., comparing paternal and maternal allele of one individual or comparing tumor and normal sample of one cancer patient. Simulation studies demonstrate the high sensitivity and specificity of IsoDOT. We apply IsoDOT to study the effects of haloperidol treatment on mouse transcriptome and identify a group of genes whose isoform usages respond to haloperidol treatment. \\

keywords: RNA-seq, isoform, penalized regression, differential isoform expression, differential isoform usage

\end{abstract}

%%%%%%%%%%%%

In the genomes of higher eukaryotes, the DNA sequence of a gene often includes multiple exons that are separated by introns. A multi-exon gene may encode several RNA isoforms and each RNA isoform includes a subset of the exons. Recent studies have shown that more than 90\% of human genes have multiple RNA isoforms which may be differentially expressed across tissues or developmental stages \citep{Wang08Alt,Pan08}, and about 75\% of human genes produce multiple RNA isoforms in a given cell type \citep{djebali2012landscape}. The study of RNA-isoform expression and its regulation is of great importance to understand the functional complexity of a living organism, the evolutionary changes in transcriptome \citep{barbosa2012evolutionary}, and the genomic basis of human diseases \citep{Wang07splicing}. \\

Gene expression is traditionally measured by microarrays. Most microarray platforms provide one measurement per gene, which does not distinguish the expression of multiple isoforms. Exon arrays can be used to study RNA isoform expression \citep{purdom2008firma,richard2010prediction}. However, RNA sequencing (RNA-seq) provides much better data for this purpose \citep{Wang2009rna}. In an RNA-seq study, fragments of RNA molecules (typically 200-500 bps long) are reverse transcribed and amplified, and then sequenced on one end (single-end sequencing) or both ends (paired-end sequencing). A sequenced end is called an RNA-seq read, which could be 30-150 bps or even longer. These RNA-seq reads are mapped to reference genome and the number of RNA-seq fragments overlapping each gene can be counted. The expression of the $j$-th gene in the $i$-th sample can be measured by normalized fragment count after adjusting for read-depth of the $i$-th sample and the length of the $j$-th gene \citep{mortazavi2008mapping}. \\

The major challenge for RNA isoform study is that we cannot directly observe the expression of each RNA isoform. More specifically, an RNA-seq fragment may be compatible with more than one RNA isoform, and thus we cannot unambiguously assign it to an RNA isoform. Several methods have been developed to address this challenge \citep{Jiang09,salzman2011statistical,richard2010prediction,xing2006expectation,Trapnell10,roberts2011improving,li2010rna,katz2010analysis,pachter2011models,chen2012statistical}. Moreover, the annotation of RNA isoforms may not be complete or accurate and thus one may need to reconstruct transcriptome annotation using RNA-seq data \citep{denoeud2008annotating}. Simultaneous transcriptome reconstruction and isoform abundance estimation can be achieved using different approaches, including penalized regression methods \citep{xia2011nsmap,bohnert2010rquant,li2011isolasso,li2011sparse}, where each possible isoform is treated as a covariate in a regression problem. Interested readers are referred to \cite{alamancos2014methods} for a comprehensive list of relevant statistical/computational methods.  \\

{Although many methods have been developed to estimate RNA isoform expression, only a few methods have been developed to assess differential isoform expression (DIE) while modeling biological variability and accounting for the uncertainty of isoform expression estimation. These methods include BitSeq \citep{glaus2012identifying}, Cuffdiff2 \citep{trapnell2013differential}, and EBseq \citep{leng2013ebseq}. All three methods are designed for two-group or multi-group comparison with multiple samples per group. BitSeq (Bayesian Inference of Transcripts from Sequencing data) adopts a two-stage approach. The first stage is isoform expression estimation within each sample using a Bayesian MCMC method. The second stage is to assess differential expression of each isoform using the posterior samples from the first stage. Cuffdiff2 employs a likelihood-based approach for isoform expression estimation and relevant hypothesis testing. For each gene, Cuffdiff2 first estimates expectation and covariance of the expression of multiple isoforms, and then uses these estimates to assess differential gene expression and differential isoform usage (DIU), where isoform usage refers to the relative expression of RNA isoforms with respect to the total expression of the corresponding gene. For differential expression, Cuffdiff2 constructs a test statistic of log fold change, standardized by its standard error. Cuffdiff2 offers two tests for differential isoform usage (DIU): for all the isoforms sharing a transcription starting site (TSS) and for differential usage of TSSs. The test statistic for DIU is the square root of the Jensen-Shannon divergence, divided by its standard error. While both BitSeq and Cuffdiff2 first estimate inform expression and then perform hypothesis testing, EBSeq uses isoform expression estimates from other methods. For two-group or multi-group comparison, EBSeq assumes the (rounded) expression estimate of an isoform follows a negative binomial distribution with group-specific mean and overdispersion. EBSeq stratifies all the RNA isoforms into multiple categories to allow category-specific mean-variance relations. These isoform categories are constructed based on the difficulty of isoform expression estimation. For example, genes with one, two, or more isoforms may form three categories. } \\

{BitSeq, Cuffdiff2, and EBseq all address an important issue for differential isoform expression (DIE):  to account for the uncertainty inherent in the isoform expression estimation process. However, there are two types of commonly encountered tasks that cannot be accomplished by these methods: to assess DIE with respect to a continuous variable, e.g., age or additive coding of genotype (i.e., 0, 1, 2, for genotype AA, AB, and BB), and to assess DIE across two groups with only one sample per group, which is not an uncommon situation in real data studies. For example, one may compare isoform expression between paternal allele and maternal allele of an individual or between normal and cancer tissues of a patient. In such situation, the RNA-seq data allow a valid statistical test, although the population for statistical inference is limited to the tested case and control (i.e., what happens if we collect more RNA-seq fragments) rather than the general case and control populations (i.e., what happens if we collect more samples from case or control population). BitSeq and EBseq cannot compare two groups with one case and one control. Cuffdiff2 provides an ad-hoc implementation for this problem. Specifically, when there is one case and one control, Cuffdiff2 estimates isoform expression variance by combining case and control, which implicitly assumes most isoforms are not differentially expressed. Therefore it is expected that Cuffdiff2 would have conservative p-value and limited power in this situation, which is confirmed in our simulation studies.} \\

{In this paper, we develop a statistical method named IsoDOT, which assesses DIE or DIU using RNA-seq data and addresses the aforementioned two tasks that cannot be accomplished by existing methods. IsoDOT treats all the RNA isoforms of a gene (or a transcript cluster of a few overlapping genes) as a unit and test whether any of these RNA isoforms is associated with the covariate of interest. Alternative strategies would be to assess differential expression or differential usage of each exon set or each RNA isoform. For testing at exon set level, the number of tests is much larger than gene-level testing, which increases the burden on multiple testing correction. In fact, multiple testing correction is also more challenging because multiple exon sets of the same gene often have correlated expression. For isoform-level testing, the major challenge is to incorporate the uncertainty in isoform expression estimation into the testing step. It is possible that two isoforms are very similar and thus available data cannot distinguish them. Therefore differential expression testing for these two isoforms separately is problematic. By performing testing per transcript cluster, IsoDOT bypasses the limitation of exon-set-level or RNA-isoform-level testing. After transcript clusters with significant DIE or DIU being identified, one may follow up on these transcript clusters to identify differentially expressed exon sets \citep{anders2012detecting} or isoforms \citep{glaus2012identifying,trapnell2013differential,leng2013ebseq}.} \\

\section*{Materials and Methods}

\subsection*{An overview}

We assume that the locations and sizes of all the exons of a gene are known. If needed, one can use existing software (e.g., TopHat \citep{Trapnell09}) to detect previously unknown exons. The inputs of our method are the bam files of all samples. From each bam files, we will derive the number of RNA-seq fragments overlapping each exon set (an exon set includes one or more adjacent or non-adjacent exons) and the distribution of RNA-seq fragments' lengths (Figure 1(a)). IsoDOT outputs the estimates of RNA isoforms' expression across all the samples, and two p-values for each gene: one for testing differential isoform expression (DIE) and one for testing differential isoform usage (DIU). The DIE test asks whether the absolute expression of any isoform of a gene is associated with the covariate of interest. In contrast, the DIU test asks, after adjusting for total expression of the corresponding gene, whether the relative expression of any isoform of this gene is associated with the covariate of interest. \\

\begin{figure}[htbp]
  \centering
  \includegraphics[width=5in]{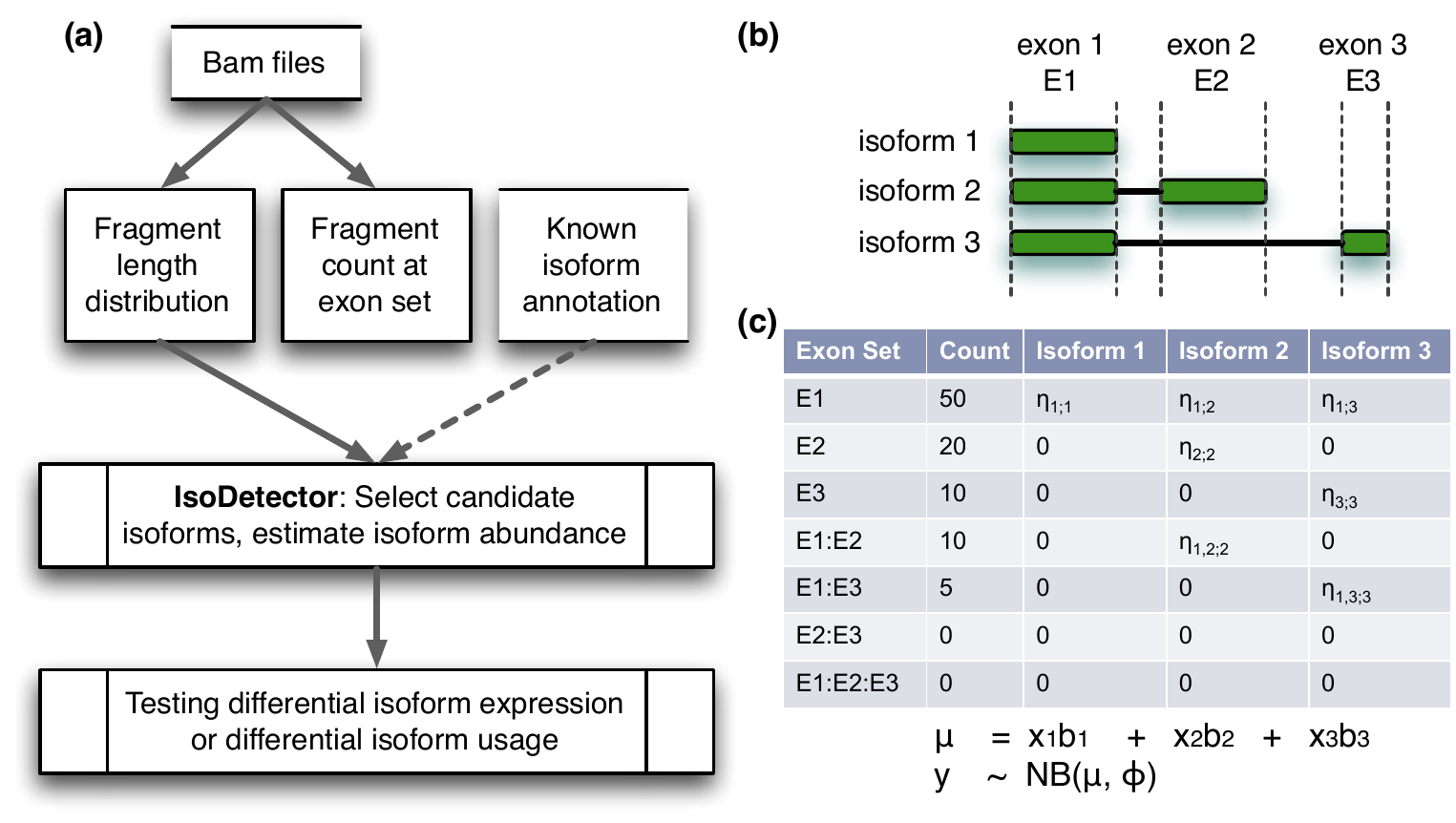}
  \caption{(a) A flow chart IsoDOT work flow. The dash line indicates know isoform annotation (i.e., transcriptome annotation) is optional. (b) A gene with 3 exons and 3 possible isoforms. (c) A matrix of input data. Each row corresponds to an exon set. The column ``Count'' is the number of RNA-seq fragments at each exon set, and the columns ``isoform $k$'' for $k=1, 2, 3$ give the effective lengths of each exon set within each isoform, and specifically, $\eta_{A,k}$ is the effective length of exon set $A$ for the $k$-th isoform. NB($\mu$,$\phi$) indicates a negative binomial distribution with mean $\mu$, and dispersion parameter $\phi$.}
\end{figure}

As part of IsoDOT, we have developed a penalized regression method, named IsoDetector, to estimate RNA isoform expression. In contrast to existing methods \citep{xia2011nsmap,bohnert2010rquant,li2011isolasso,li2011sparse}, IsoDetector employs penalized negative binomial regression with a log penalty. {The negative binomial distribution is commonly used to model RNA-seq data, and previous studies have shown that negative binomial distribution is able to account for the variation of RNA-seq fragment counts across biological replicates \citep{Langmead10}. Many popular methods for differential expression testing, such as DEseq \citep{anders2010differential} and edgeR \citep{robinson2010edger}, adopt negative binomial distribution assumption. More specifically, the negative binomial distribution assumption, denoted by $NB(\mu, \phi)$, implies that the RNA-seq fragment count across biological replicates follows a negative binomial distribution with mean value $\mu$ and variance $\mu + \mu^2 \phi$, where $\phi$ is an over-dispersion parameter. Therefore, the variance of a negative binomial distribution can be arbitrarily large for a large value of $\phi$.} The Log penalty, which can be interpreted as iterative adaptive Lasso penalty \citep{Tibshirani96,zou2006adaptive,Sun10}, is flexible enough to handle a broad class of penalization problems \citep{Chen2013}. IsoDOT can test DIE/DIU against any categorical or continuous covariate at any sample size, with or without known isoform annotation. Our simulation and real data analysis demonstrate the satisfactory performance of IsoDOT for RNA-seq data from human or mouse, while IsoDOT can be applied to analyze RNA-seq data of any species that has a reference genome. \\

Two exons of a gene may overlap partially. In such situation, we split them into three exons: the part that only belongs to the first or the second exon, and the part that belongs to both exons. Multiple genes may overlap on one or more exons, and we consider these genes as a transcript cluster. We further impose a constraint that an isoform is unlikely if any two exons of this isoform are unlikely to belong to the same gene. 

\subsection*{Isoform estimation in a single sample.} We study the isoforms of each transcript cluster separately, and the following discussions apply to a specific transcript cluster. Denote the number of exons of a transcript cluster by $k$. Let $A$ be an exon set, i.e., a subset of the $k$ exons.  Let $y_{iA}$ be the number of sequence fragments that overlap and only overlap with all the exons of $A$ in the $i$-th sample, where $1 \leq i \leq n$, and $n$ is the sample size. A sequence fragment overlaps with an exon if the ``sequenced portion'' of this fragment overlaps with $\geq$ 1 bp of the exon. For example, if a fragment is sequenced by a paired-end read where the first end overlaps with exon 1 and 2 and the second end overlaps with exon 4, then this fragment is assigned to exon set $A = \{1, 2, 4\}$. \\

To illustrate the main feature of our method, we consider a gene (which is a transcript cluster itself) with 3 exons and 3 isoforms (Figure 1(b)). Denote its expression at sample $i$ by $\by_i = (y_{i\{1\}}, y_{i\{2\}}, y_{i\{3\}}, y_{i\{1,2\}}, y_{i\{2,3\}}, y_{i\{1,3\}}, y_{i\{1,2,3\}})\trans$. We assume $y_{iA}$ follows a negative binomial distribution $\psi(\mu_{iA}, \phi)$ with unknown mean $\mu_{iA}$ and dispersion parameter $\phi$. Let $\boldsymbol{\mu}_i$ be a column vector concatenating the $\mu_{iA}$'s, then $\boldsymbol{\mu}_i = E(\by_i)$. We model $\boldsymbol{\mu}_i$ by:
\begin{eqnarray}
\boldsymbol{\mu}_i = \X_i \boldsymbol{\beta} = \sum_{u=1}^p \x_{iu} \beta_u, \ \ \ \ \beta_u \geq 0,
\label{eq:link1}
\end{eqnarray}
where $\beta_u$ is proportional to the transcript abundance of the $u$-th isoform, $\X_i = (\x_{i1}, ..., \x_{ip})$, and $\x_{iu}$ for $1 \leq u \leq p$ represents the effective lengths of all the exon sets for the $u$-th isoform in the $i$-th sample. Intuitively, \textbf{effective length} is the ``usable length'' for the data generation mechanism, i.e., the number of positions where a randomly selected sequence fragment can be sampled. The effective length of an exon set varies across the underlying isoforms. For example, the isoform 1 and 3 of the gene shown in Figure 1(b-c) do not include exon 2, and thus the effective length of exon set \{2\} is 0 for isoform 1 or 3. In contrast, the effective length of exon 2 is nonzero for isoform 2, which includes exon 2. In addition, effective length is also a function of sample-specific RNA-seq fragment length distribution. The typical length of RNA-seq fragments is often chosen in the RNA-seq library preparation. However the fragment length distribution is unknown and can be estimated from observed RNA-seq data. See Supplementary Materials Section A for details. In this example, the design matrix includes the effective lengths of all the exon sets for isoforms 1, 2, and 3. Next, we recast the isoform estimation problem to a negative binomial regression problem with fragment counts $\y_i$ as response and effective lengths $\X_i$ as covariates: 
\begin{eqnarray}
\y_i \sim \psi(\boldsymbol{\mu}_i, \phi), \textrm{ and }
\boldsymbol{\mu}_i = \X_i \boldsymbol{\beta}.
\label{eq:mod2}
\end{eqnarray}
Equation (\ref{eq:mod2}) should be understood such that $y_{iA}$'s are independent with each other and $y_{iA}$ follows a negative binomial distribution $\psi(\mu_{iA}, \phi)$. The independence assumption is reasonable because the the fragment counts across exon sets should be independent given isoform configurations. \\

The regression problem presented in equation (\ref{eq:mod2}) is challenging because there can be a large number of possible isoforms and their effective lengths (e.g., the columns of the design matrix $\X_i$) may be linearly dependent or significantly correlated. To address this difficulty, we first select a set of candidate isoforms, and then apply a penalized negative binomial regression to select the final set of isoforms from these candidate isoforms. The candidate isoforms can be selected using observed RNA-seq data (Supplementary Materials Section B) or transcriptome annotation database (e.g., Ensembl \citep{Flicek11}). \\

{In our analysis, we skip the exon sets that have zero or negligible effective lengths across all the candidate isoforms because these exon sets are not informative for isoform expression estimation. For example, the exon set $\{2,3\}$ or $\{1,2,3\}$ in the example shown in Figure 1 (c) are not included in the analysis.} The number of candidate isoforms, denoted by $p$, can be much larger than the number of (informative) exon sets, denoted by $m$, and there may be high correlations among the effective lengths of the candidate isoforms. Therefore, selecting of a final set of isoforms from the candidate isoforms is a challenging variable selection problem. Lasso penalty has been applied in previous studies. However, the selection consistency of Lasso requires an \textit{irrepresentability condition} on the design matrix \citep{zou2006adaptive,zhao2006model}, which posits that there are weak correlations between the ``important covariates'', which have non-zero effects and the ``unimportant covariates'', which have zero effects. This irrepresentability condition is often not satisfied for isoform selection problem due to the high correlations among candidate isoforms. We employ a Log penalty \citep{mazumder2011sparsenet} for this challenging variable selection problem, which does not require irrepresentability condition and can be interpreted as iterative adaptive Lasso \citep{Sun10,Chen2013}. The algorithm for fitting this penalized negative binomial regression is outlined in Supplementary Materials Section C.

\subsection*{Isoform estimation in multiple samples.}
To estimate isoform expression in multiple samples, we have to account for read-depth difference across samples. Let $t_{i}$ be a read-depth measurement for the $i$-th sample. For example, $t_i$ can be the total number of RNA-seq fragments in the $i$-th sample, or the 75 percentile of the number of RNA-seq fragments per gene in the $i$-th sample \citep{bullard2010evaluation}. We first consider the case without any covariate associated with isoform expression. To account for read-depth variation, we modify equation (\ref{eq:link1}) to 
\begin{eqnarray}
\boldsymbol{\mu}_i = t_{i} \X_i \boldsymbol{\gamma} = \sum_{u=1}^p t_{i}\x_{iu} \gamma_u,
\label{eq:link2}
\end{eqnarray}
where $\gamma_u$ is proportional to relative expression of the $u$-th isoform, after normalizing by $t_{i}$. \\

Let $\y\trans = (\y_1\trans, ..., \y_n\trans)$,
$\boldsymbol{\mu}\trans = (\boldsymbol{\mu}_1\trans, ..., \boldsymbol{\mu}_n\trans)$, and
$\Z\trans=(t_{1}\X_1\trans, ..., t_{n} \X_n\trans)$, where $\y$ and $\boldsymbol{\mu}$ are vectors of length $nm$ and $\Z$ is a matrix of size $nm \times p$. Recall that $p$ is the number of candidate isoforms, $n$ is sample size, and $m$ is the total number of exon sets. Then the isoform selection problem can be written as a negative binomial regression problem
\begin{eqnarray}
\y \sim \psi(\boldsymbol{\mu}, \phi) \textrm{ and }
\boldsymbol{\mu} = \Z \boldsymbol{\gamma}, \quad
\textrm{where $\gamma_j \geq 0$ for $1 \leq j \leq p$.}
\label{eq:mod4}
\end{eqnarray}
After imposing penalty, we solve this regression problem using the method described in Supplementary Materials Section C.  \\

Next we consider isoform estimation given a continuous covariate $\g=(g_{1}, ..., g_{n})\trans$. We assume the expression of an isoform $u$ for sample $i$, denoted by $\gamma_{iu}$, is a linear function of covariate $g_i$: $\gamma_{iu} = a_u + \tilde{b}_u g_{i}$.  This linear model is an appropriate choice when $g_i$ represents SNP genotype \citep{sun2012statistical}, which is the focus of our empirical data analysis. In this linear model setup, a complex set of constraints is needed for $a_u$ and $\tilde{b}_u$ so that $\gamma_{iu} \geq 0$ for any value of $g_i$. Therefore we reformulate the problem as follows. Without loss of generality, we scale the value of $g_i$ to be within the range of [0,1] with the minimum and maximum values being 0 and 1, respectively. For example, if $g_i$ corresponds to a SNP with additive effect, we can set $g_{i}$ = 0, 0.5, or 1 for genotype AA, AB, or BB. Let $b_u = \tilde{b}_u + a_u$, then $\gamma_{iu} = a_u + (b_{u} - a_{u})g_{i} = a_u(1 - g_{i}) + b_{u} g_{i}$. Under this model, we have 
$$ 
\gamma_{iu} \geq 0 \textrm{ for any $g_i \in [0,1]$ } \Leftrightarrow a_u \geq 0 \textrm{ and } b_u \geq 0.
$$ 
Let $\ba = (a_1, ..., a_p)\trans$ and $\bb = (b_{1}, ..., b_{p})\trans$, we have 
\begin{eqnarray}
\boldsymbol{\mu}_i = t_i\X_i [\mathbf{a}(1 - g_i) + \mathbf{b} g_i].
\label{eqn:link:num}
\end{eqnarray}
By concatenating $\mathbf{a}$ and $\mathbf{b}$ into a vector: $\boldsymbol \theta =  (a_1, ..., a_p, b_1, ..., b_p)\trans$, we can rewrite equation (\ref{eqn:link:num}) as $\boldsymbol{\mu}_i =\W_i \boldsymbol \theta$,
where $\W_i  = [t_i \X_i(1 - g_i), t_i \X_i g_i]$ is an $m \times 2p$ matrix. Let $\W\trans=(\W_1\trans, ..., \W_n\trans)$, then the isoform expression estimation problem reduces to a negative binomial regression problem with non-negative coefficients 
\begin{eqnarray}
\y \sim \psi(\boldsymbol{\mu}, \phi) \textrm{ and }
\boldsymbol{\mu} = \W \boldsymbol{\theta}, \quad
\textrm{where $\theta_j \geq 0$ for $1 \leq j \leq 2p$.}
\label{eq:mod6}
\end{eqnarray}
After imposing penalty, we solve the resulting penalized regression problem by the coordinate ascend method described in Supplementary Materials Section C. \\

Finally we consider the general situation with $q$ covariates, denoted by $\g_1, ..., \g_q$, where $\g_v=(g_{1v}, ..., g_{nv})\trans$ for $v=1, ..., q$. Without loss of generality, we assume $0 \leq g_{iv} \leq 1$ for $1 \leq i \leq n$ and $1 \leq v \leq q$. Then we model $\gamma_{iu}$ by $\gamma_{iu} = qa_u + \sum_{v=1}^q(b_{vu} - a_{u})g_{iv} = a_u \sum_{v=1}^q (1 - g_{iv}) + \sum_{v=1}^q b_{uv} g_{iv}$. This is simply a multiple linear regression model where each covariate $g_v$ has its own effect. Let $\ba = (a_1, ..., a_p)\trans$ and $\bb_v = (b_{1v}, ..., b_{pv})\trans$, then
\begin{eqnarray}
\boldsymbol{\mu}_i = t_i\X_i \left[\ba \sum_{v=1}^q (1 - g_{iv}) + \bb_1 g_{i1} + \cdots + \bb_q g_{iq} \right].
\label{eqn:link:num1}
\end{eqnarray}
By concatenating $\ba$, $\bb_1$, $\bb_2$, ..., $\bb_q$ into a vector: $\boldsymbol \theta =  (\ba\trans, \bb_{1}\trans, ..., \bb_{q}\trans)\trans$, we rewrite the above equation as $\boldsymbol{\mu}_i =\W_i \boldsymbol \theta$,
where $\W_i  = [t_i \X_i \sum_{v=1}^q (1 - g_{iv}), t_i \X_i g_{i1}, \cdots, t_i \X_i g_{iq}]$ is an $m \times (q+1)p$ matrix. Let $\W\trans=(\W_1\trans, ..., \W_n\trans)$, then we form an negative binomial regression problem
\begin{eqnarray}
\y \sim \psi(\boldsymbol{\mu}, \phi) \textrm{ and }
\boldsymbol{\mu} = \W \boldsymbol{\theta}, \quad
\textrm{for $\theta_j \geq 0$, $1 \leq j \leq (q+1)p$.}
\label{eq:mod8}
\end{eqnarray}
After imposing penalty on each $\theta_j$, we can solve the resulting penalized regression problem by the coordinate ascend method described in Supplementary Materials Section C. \\

When studying multiple samples, it is possible that an exon set is only expressed in a subset of samples so that when examining the fragment counts of this exon set across all samples, there are more 0's than expected by a negative binomial distribution. In such case, we introduce a zero-inflated component and employ the zero-inflated negative binomial distribution \citep{rashid2011zinba} to model RNA-seq fragment count data. 

\subsection*{Testing for differential isoform expression.}
We have described the statistical model to estimate RNA isofrom expression in multiple samples given one or more covariate. {Build on this model, we assess differential isoform expression with respect to a set of covariates denoted by $V$ using a likelihood ratio test. Specifically, the null hypothesis ($H_0$) is that $b_{uv}=a_u$ for $u=1, ..., p$ and $v \in V$ and the alternative hypothesis ($H_1$) is that $b_{uv} \neq a_u$ for at least one pair of ($u$, $v$), where $u=1, ..., p$ and $v \in V$.} It is helpful to understand this test by considering two special cases. First, we assume there is only one numerical covariate. {Under $H_0$, we solve the isoform estimation problem by a penalized negative binomial regression with expected value $\boldsymbol{\mu}_i = t_i\X_i \mathbf{a}$. Under alternative, $\boldsymbol{\mu}_i = t_i\X_i [\mathbf{a}(1 - g_i) + \mathbf{b} g_i]$. Therefore the number of parameters is $p$ under $H_0$ and $2p$ under $H_1$. The asymptotic Chi-square distribution with degree of freedom $p$ does not apply because the models are estimated, under both $H_0$ and $H_1$, by penalized regression. In the second special case, we assume the only variable of interest is a categorical variable with $d$ categories. This categorical variable can be coded as $d-1$ binary variables, dented by $g_{i1}, ..., g_{i,d-1}$. The expected values of fragment counts across exon sets under $H_0$ and $H_1$ are $\boldsymbol{\mu}_i = (d-1) t_i\X_i \mathbf{a}$ and $\boldsymbol{\mu}_i = t_i\X_i [\ba \sum_{v=1}^{d-1} (1 - g_{iv}) + \bb_1 g_{i1} + \cdots + \bb_{d-1} g_{i,d-1}]$, respectively. Therefore the number of parameters under $H_0$ and $H_1$ are $p$ and $dp$, respectively. Again, the asymptotic Chi-square distribution with degree of freedom $(d-1)p$ does not apply because the models are estimated, under both $H_0$ and $H_1$, by penalized regression.} \\

We use likelihood ratio (LR) statistic as our test statistic: 
\begin{eqnarray}
\mathcal{LR}=2(\ell_1-\ell_0),
\end{eqnarray}
where $\ell_0$ and $\ell_1$ are the log likelihoods under null and alternative, respectively. 
Because of penalized estimation, the null distribution of this test statistic no longer follows the standard asymptotic distribution. We obtain the null distribution by parametric bootstrap or permutation. The parametric bootstrap approach proceeds as follows. (1) Fit the penalized negative binomial regression under null. (2) Sampling fragment counts based on the fitted null model. (3) Using the sampled counts to refit models under null and alternative and calculate the LR statistic. (4) Repeat the steps (1)-(3) a large number of times as needed \citep{jiang2012statistical}. (5) Calculate the p-value as the proportion of the bootstrapped LR statistics that are larger than $\mathcal{LR}$. This parametric bootstrap procedure yields valid p-values regardless of the sample size $n$. However, since the sampling population is all the RNA-seq fragments from the studied samples, small p-values only imply significant difference of the studied samples and should not be generalized to other individuals. \\

If sample size is sufficiently large (e.g., $\geq$ 5 cases vs. $\geq$ 5 controls), we can obtain valid statistical inference for the corresponding population (instead of studied samples) by calculating permutation p-values. Specifically, the null model is fitted without the covariate of interest, and thus its log likelihood $l_0$ remains unchanged across permutations. In each permutation, we permute the covariate of interest and refit the alternative model, and then calculate the likelihood ratio test statistic. We repeat this process a large number of times to obtain null distribution of the likelihood ratio statistic.  

\subsection*{Testing for differential isoform usage.}
All the previous discussions, including RNA isoform expression estimation and differential isoform expression testing, focus on absolute expression of RNA isoforms, which is RNA isoform expression after correcting for read-depth variation across samples. Another measure of RNA isoform expression, which may be more interesting in many situations, is the relative expression with respect to the total expression of the corresponding gene or transcript cluster.  This is because a gene may have higher or lower expression overall, and it may also switch the usage of some RNA isoforms. For example, a gene may predominately use one isoform in one tissue and switch to another isoform in the other tissue. We use the word isoform usage to refer to the relative expression of an RNA isoform. To study isoform usage in the $i$-th sample, we condition on the total number of RNA-seq fragments of the corresponding transcript cluster in the $i$-th sample, denoted by $r_{i}$. To study RNA isoform usage, we just need to replace $t_i$ in the aforementioned models with $r_i$. \\

\section*{Results and Discussions}

\subsection*{Simulation for case-control comparison.}\ \ 
We simulated $\sim$1 million 76 + 76 bps paired-end RNA-seq reads for a single case and control sample respectively using Flux Simulator (\url{http://flux.sammeth.net/simulator.html}) and the Ensembl transcriptome annotations (version 67, \url{http://useast.ensembl.org/info/data/ftp/index.html}) for chromosome 1 (chr1) and chromosome 18 (chr18) for the mouse genome. We simulated the data such that all the genes from chr1 were equivalently expressed and all the genes from chr18 were differentially expressed, either in terms of total expression or isoform usage (Supplementary Figure 2). These simulated RNA-seq reads were mapped to the reference genome using Tophat \citep{Trapnell09}. Next, we counted the number of RNA-seq fragments per exon set. It was important to consider all exon sets rather than just exon or exon junctions because many RNA-seq fragments overlap with more than two exons. For example, in this simulated dataset, $\sim$27\% of the paired-end reads overlapped 3 or more exons (Supplementary Figure 3).  In addition, we confirmed that the number of sequence fragments per exon set was proportional to the effective length of the exon set (Supplementary Figure 4).\\

The dimension of this problem (i.e., the number of exon sets $m$ versus the number of candidate isoforms $p$) was illustrated in Supplementary Figures 5-6. Given transcriptome annotation, $p < m$ for the vast majority of transcript clusters, and without transcriptome annotation, we restricted the number of candidate isoforms so that approximately $p < 10m$. In either case, there were strong correlations among the candidate isoforms (Supplementary Figures 7-8), therefore necessitating the use of penalized regression in IsoDetector. After applying penalized regression, most transcription clusters included 10 or fewer isoforms, with or without transcriptome annotation (Supplementary Figures 9-10). \\

We compared isoform abundance estimates from IsoDetector and Cufflinks (v2.0.0) \citep{Trapnell10,trapnell2013differential} using the case sample. The conclusions from the control sample were similar (results not shown). We focused on 1,062 transcript clusters (corresponding to 5,185 transcripts) that had at least 2 exons with $\geq$5 sequence fragments overlapping each exon. Most of these transcript clusters included 1-2 genes and 1-42 transcripts, and most of the 5,185 transcripts harbored 6-500 RNA-seq fragments (Figure 2a). 
\begin{figure}[htbp]
  \centering
\includegraphics[width=3.5in]{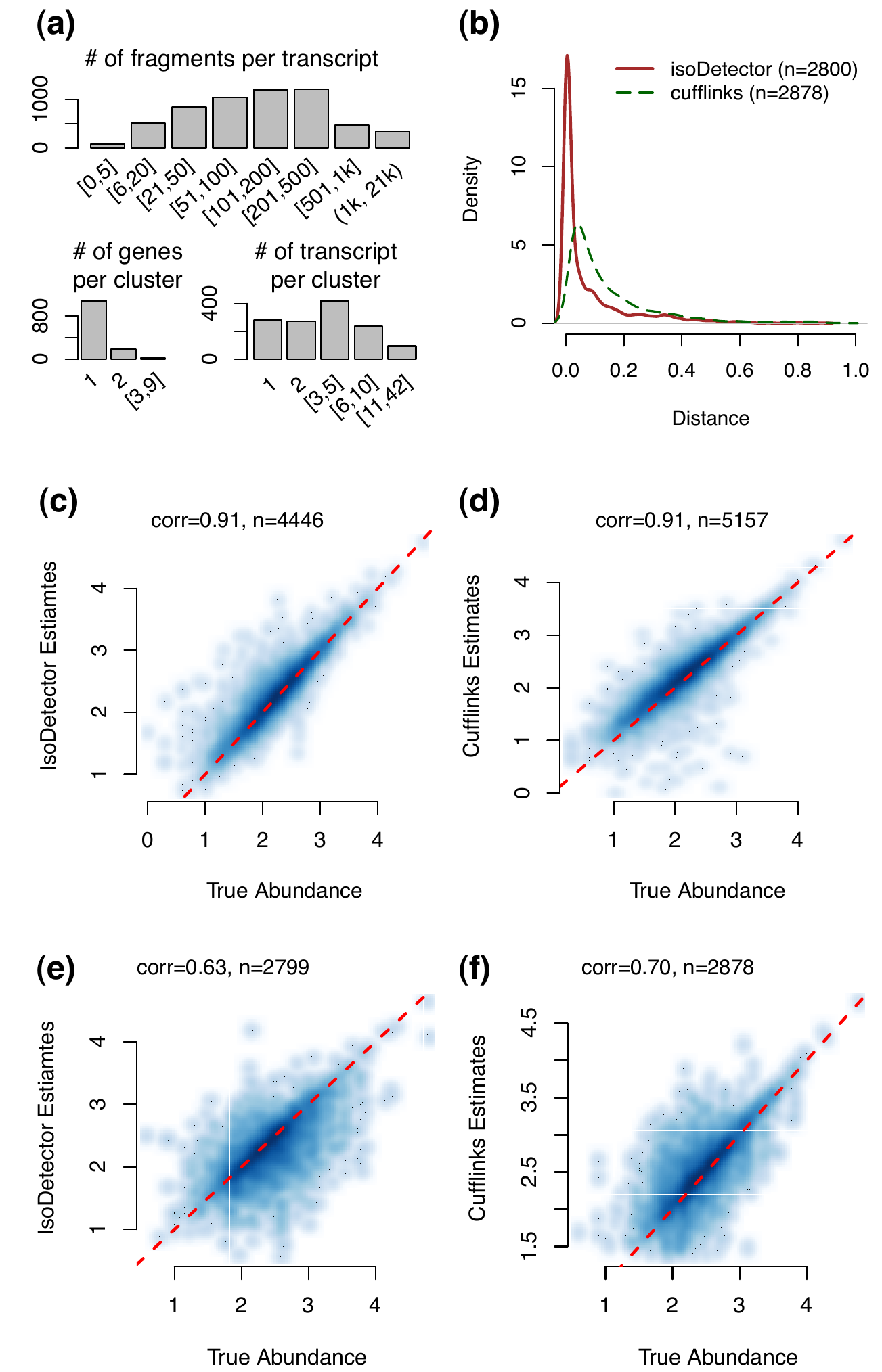}
\caption{(a) A summary of the RNA-seq data and annotation of  the simulated case sample. 
(d) Density curves of the distance between each de novo transcript and its closest transcript from transcriptome annotation. The distance is defined as the ratio of the number of unmatched base pairs over the total number of base pairs covered by either isoform. A base pair is ``matched'' if it corresponds to an exon or intron location for both isoforms.
(c-d) Comparison of true transcript abundance vs. the estimates by IsoDetector or the Cufflinks when we use known isoform annotation. Both X and Y-axes are in $\log_{10}$ scale. ``n'' is the number of transcripts with status ``OK'' for either IsoDetector or Cufflinks.  (e-f) Comparison of true transcript abundance vs. the estimates by IsoDetector or Cufflinks when we do not use any isoform annotation. }
\end{figure}
If transcriptome annotation was unavailable, the isoforms selected by IsoDetector were more similar to the true ones than Cufflinks (Figure 2b). The two methods had similar accuracy in terms of transcript abundance estimation, either with (Figure 2c-2d) or without (Figure 2e-2f) transcriptome annotation. \\

We next compared the power of IsoDOT to that of Cuffdiff (v2.0.0) \citep{trapnell2013differential} in terms of testing for differential expression or differential isoform usage. The results of Cuffdiff were obtained from three files: \texttt{gene\_exp.diff} (differential expression), \texttt{splicing.diff} (differential usage of the isoforms sharing a transcription start site (TSS)), and \texttt{promoters.diff} (differential usage of TSSs). Majority of the genes in file \texttt{gene\_exp.diff} have status ``OK'' and they were used in the following comparison. However, no gene has status ``OK'' in file \texttt{splicing.diff}  or \texttt{promoters.diff}. In these two files, all the genes with meaningful p-values have status ``NOTEST'', indicating that Cuffdiff recommends the users not to trust these testing results. The reason is as follows. For two group comparison with one case and one control, Cuffdiff combines the case and the control to estimate variance of biological replicates, which leads to very conservative p-value since this approach implicitly assumes the case and the control have the same expected value. Nonetheless, these genes with status ``NOTEST'' were used in the following comparison because they are the only genes that we can use. A gene could have multiple p-values in \texttt{splicing.diff}. In favor of Cuffdiff for power comparison, we used the smallest p-value for each gene. Based on our simulation setup, power was defined as the proportion of the genes from chr18 that had significant DIE or DIU. IsoDOT had substantially higher power than Cuffdiff (Figure 3), 
\begin{figure}[htbp]
  \centering
\includegraphics[width=4in]{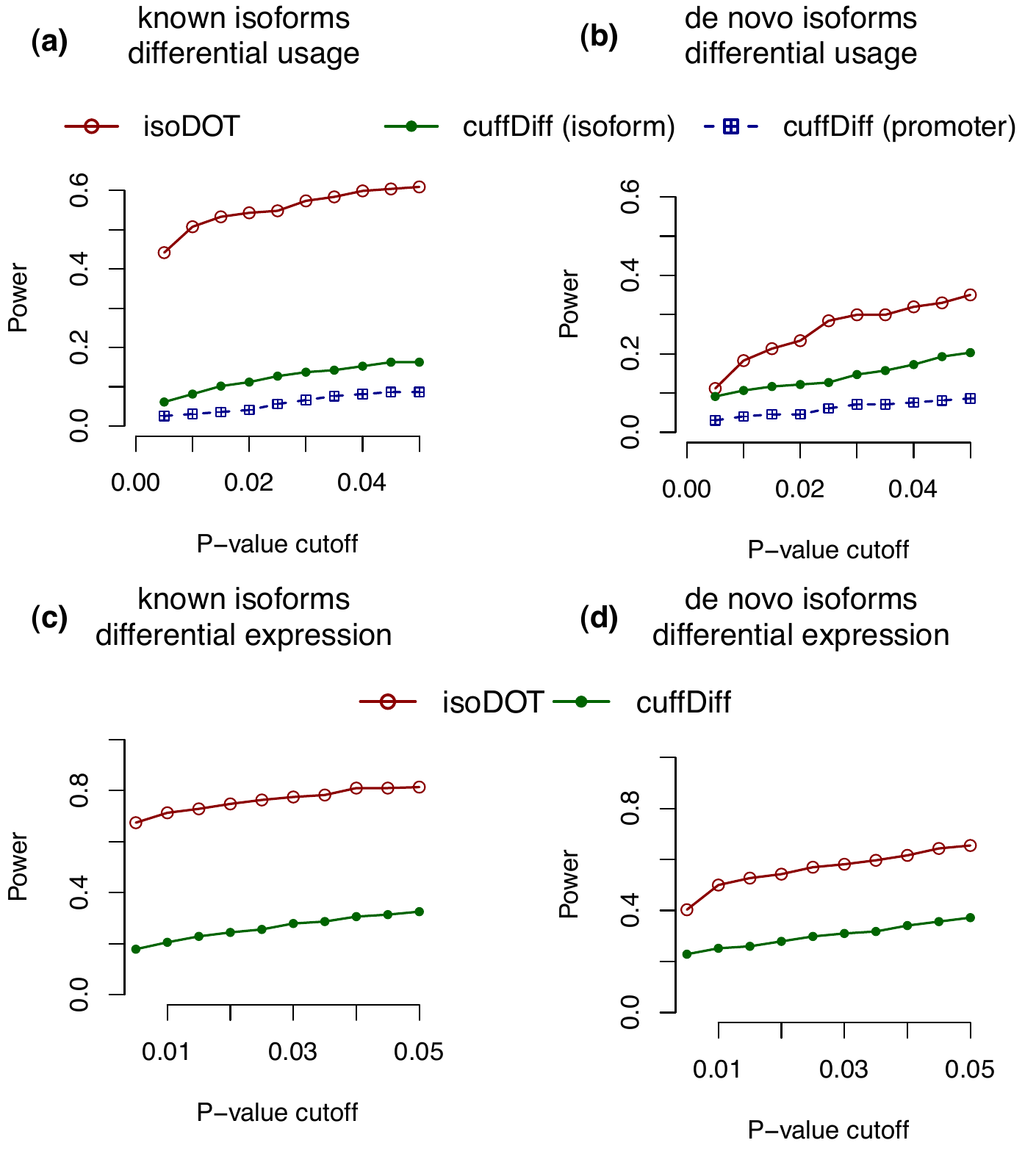}
\caption{Compare the power of IsoDOT and Cuffdiff for detesting genes with differential isoform usage (a-b) or differential isoform expression (c-d), while transcriptome annotation is known (a,c) or not (b,d). }
\end{figure}
attributed to the correct type I error control of IsoDOT (Supplementary Figure 11). {The poor performance of Cuffdiff is not simply a calibration issue because IsoDOT still performs better than Cuffdiff when we compare two methods using ROC curves (Supplementary Figure 12). Instead, poor performance of Cuffdiff is because it tries to estimate variance across biological replicates when there is no biological replicate at all. IsoDOT can perform a valid test because it does not try to estimate variance of biological replicates. Instead, it tries to estimate the variance due to resampling of RNA-seq reads. To be fair, Cuffdiff does recommend such test with one case versus one control and we include Cuffdiff in our comparison because it is the only method that allows for such testing.} In conclusion, this simulation illustrated that IsoDOT worked well in this challenging situation of comparing one case vs. one control. 
 
 \subsection*{Simulation for isoform usage eQTL.}
To illustrate the performance of IsoDOT in testing differential isoform usage with respect to a continuous covariate, we applied IsoDOT to map isoform usage eQTL (gene expression quantitative trait locus) using simulated RNA-seq data and real genotype data. We downloaded genotype data from 60 European HapMap samples \citep{Thorisson2005} and selected 949,537 SNPs with minor allele frequency (MAF) $>$ 0.05 for the following analysis. We defined transcription clusters based on Ensembl annotation (version 66, \url{http://useast.ensembl.org/info/data/ftp/index.html}), and selected 200 transcript clusters that satisfied the following two conditions for our simulation studies. (1) Each transcript cluster has $>$ 1 annotated RNA-isoforms, and (2) Each transcript cluster has $\geq$ 1 SNP on the gene body (any intronic or exonic regions) or within 1000bp of the gene body. For each selected transcript cluster, we simulated RNA-seq fragment counts across exon sets under null ($H_0$) and alternative ($H_1$), respectively. Specifically, for each gene, we randomly selected 50\% the isoforms to have zero expression and set the expression of the other 50\% of isoforms by drawing $\gamma_u$ (equation~(\ref{eq:mod4})) from a uniform distribution U[0.5, 1].  \\

Then using these simulated data, we assessed the differential isoform usage of each transcript cluster with respect to each of the nearby SNPs (within 1000bp of the gene body) and kept the most significant eQTL per transcript cluster. For each transcript cluster, up to 1000 permutations were carried out to correct for multiple testing across the multiple nearby SNPs. Under $H_0$, the permutation p-values followed uniform distribution; and under $H_1$, the permutation p-values were obviously enriched with small values (Figure 4(a)). 
\begin{figure}[htbp]
  \centering
\includegraphics[width=4in]{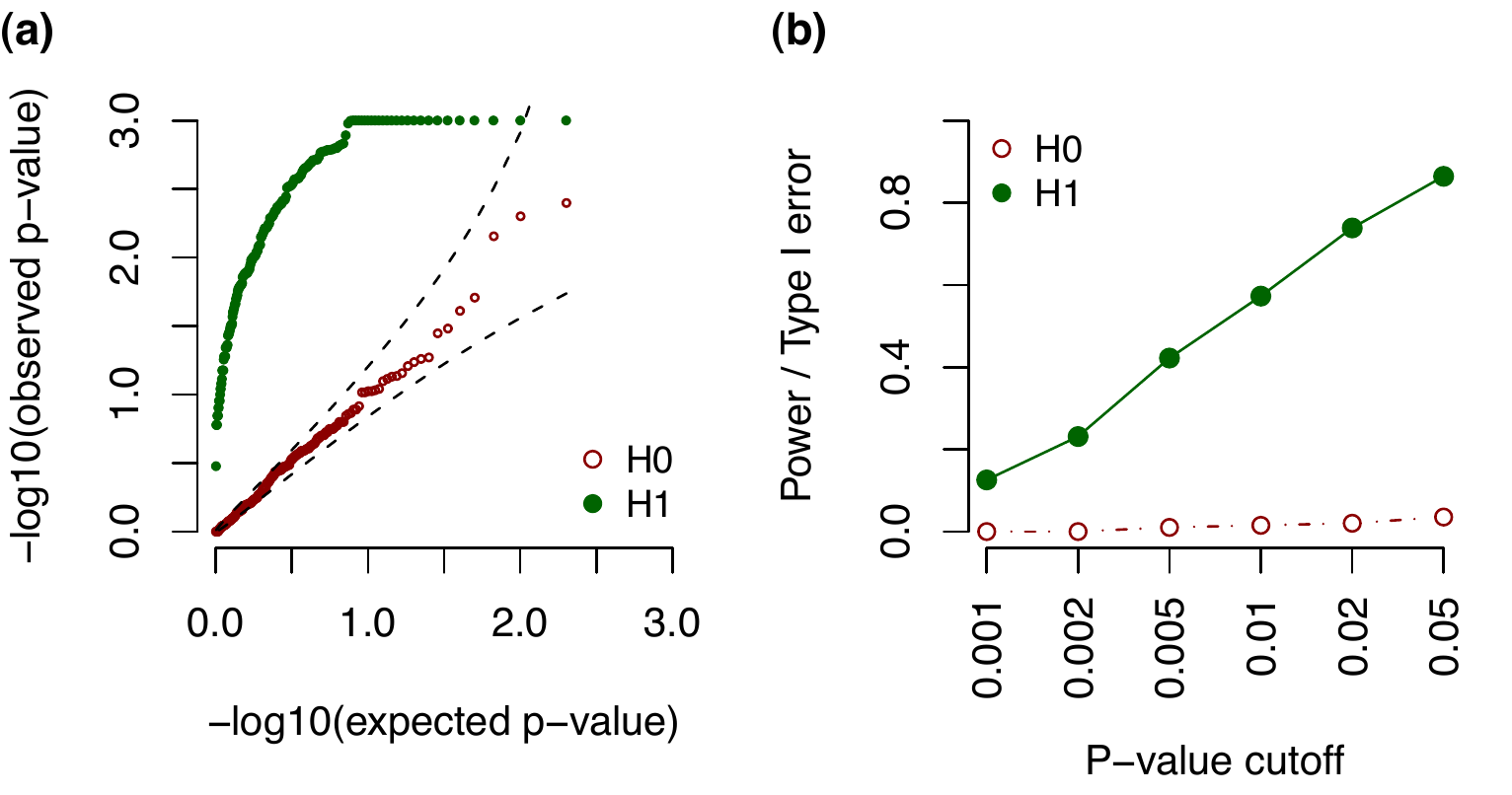}
\caption{Simulation results for testing isoform-usage eQTL on 200 transcription clusters.  (a) The qq-plot for p-values distributions against expected uniform distribution. (b) Power (H1) or type I error (H0) for different p-value cutoffs. }
\end{figure}
In this simulation, we had $\sim$ 80\%/40\% power to detect local isoform usage eQTL for permutation p-value cutoffs 0.05 and 0.005, respectively (Figure 4(b)). Therefore this simulation demonstrate that IsoDOT correctly control type I error and has power to detect differential isoform usage with respect to a continuous covariate.

\subsection*{Haloperdiol treatment effect on mouse transcriptome.}
Haloperidol is a drug to treat schizophrenia, acute psychotic states, and delirium. An adverse side effect of chronic haloperidol treatment is tardive dyskinesia (TD). Haloperidol-induced vacuous chewing movements (VCMs) in mice is a valid animal model for TD \citep{crowley2012antipsychotic}. We applied IsoDOT to analyze RNA-seq data for mice treated with haloperidol vs. placebo with particular interest in identifying genes responding to haloperidol treatment and/or responsible for VCM. RNA-seq data were collected from whole brains from four mice: two C57BL/6J mice treated with haloperidol or placebo and two 129S1Sv/ImJ$\times$PWK/PhJ F1 mice treated with haloperidol or placebo. Each RNA-seq fragment was sequenced on both ends by 93 or 100bp, and 20-27 million RNA-seq reads were collected for each mouse (Supplementary Table 1). See Supplementary Materials Section D for additional details of the experiment. \\

We first studied differential isoform expression/usage between two C57BL/6J mice. RNA-seq reads were mapped to the \texttt{mm9} reference genome using TopHat \citep{Trapnell09}. At FDR of 5\%,  IsoDOT identified 86 or 88 genes with differential isoform usage (DIU), with or without transcriptome annotation, respectively. For the test of differential isoform expression (DIE), also at FDR of 5\%, IsoDOT identified 332 or 206 genes with or without transcriptome annotation, respectively. We sought to gain some insight of these four gene lists by applying functional category  enrichment analysis \citep{sherman2009systematic} on the top 100 genes in each list. Only those DIU genes identified with transcriptome annotation were significantly associated with biologically relevant categories such as neuron projection (Supplementary Figure 13-16). This implied that DIU, rather than DIE, might be more relevant to haloperidol treatment in C57BL/6J mice. The gene lists identified without transcriptional annotation did not show functional enrichment, which implied larger sample size or higher read-depth were needed to detect DIE or DIU without transcriptome annotation. {According to Cuffdiff manual, it is not recommended to run Cuffdiff with sample size one versus one. However, we still evaluated the performance of Cuffidff in this dataset for comparison purpose. Using the results reported by Cuffdiff, no gene has q-value smaller than 0.05 (with or without annotation, alternative promoter or alternative splicing), and functional category enrichment analysis \citep{sherman2009systematic} on the top 100 genes reported by Cuffdiff identified no significantly enriched functional category (Supplementary Figure 17).}\\ 

{In the following, we discuss in more details several DIU genes identified by IsoDOT, which can be potential targets for follow up studies (Supplementary Table 2).} For example, \textit{Utrn} (utrophin, Figure 5) 
\begin{figure}[htbp]
  \centering
\includegraphics[width=4in]{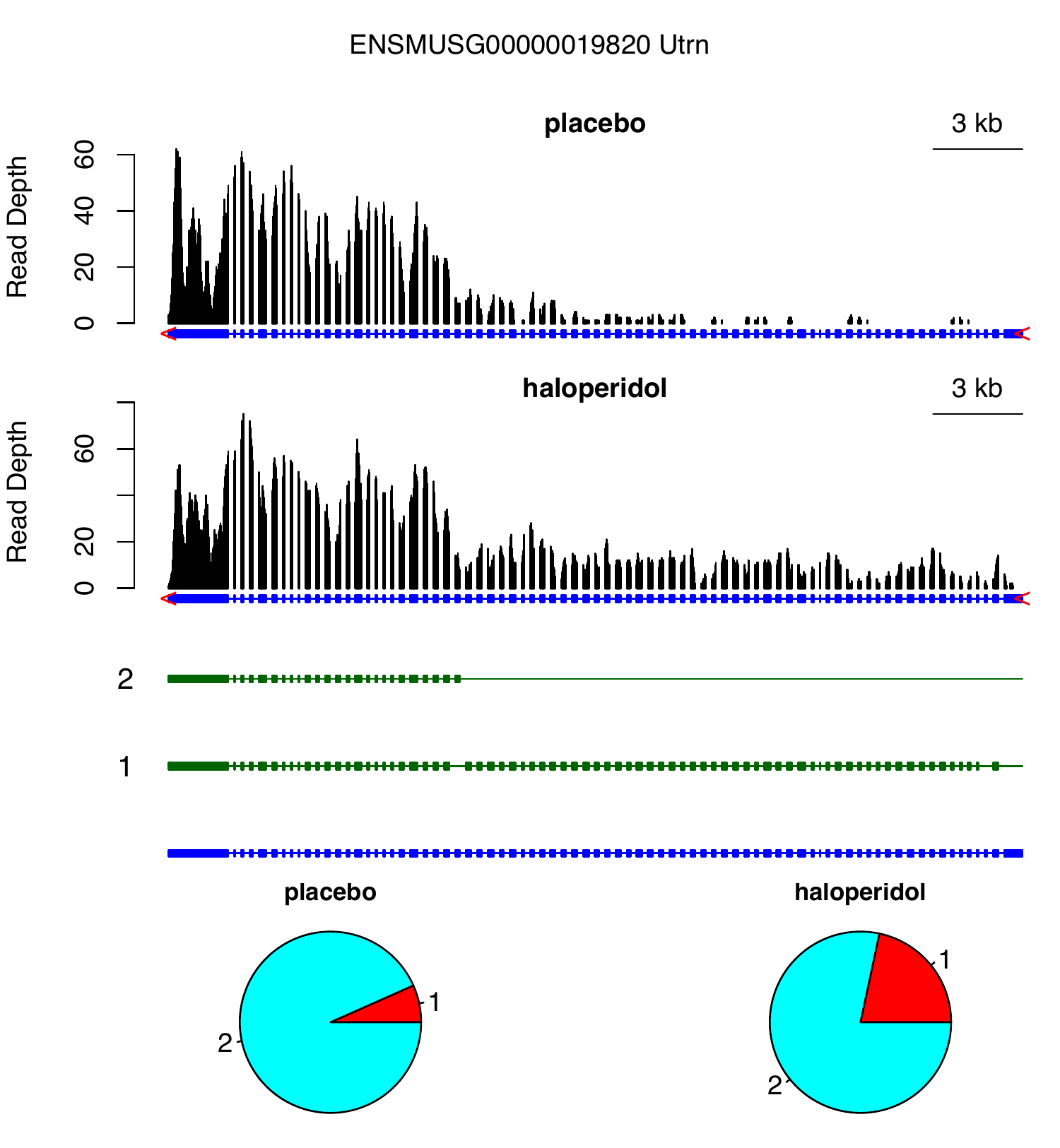}
\caption{Differential isoform usage of gene Utrn between a mouse with haloperidol treatment and a mouse with placebo. The top panel shows the read-depth of RNA-seq reads in two conditions. Two annotated isoforms and all the exons are illustrated in middle panel. The expression of each isoform under two conditions are shown in the bottom panel.}
\end{figure}
and  \textit{Dmd} (dystrophin) are both involved in neuron projection and could be candidates underlying the VCM side effect. \textit{Grin2b} (glutamate receptor, ionotropic, NMDA2B, Supplementary Figure 18) or its human ortholog is involved in Alzheimer's disease, Huntington's disease, and amyotrophic lateral sclerosis (ALS), which are potentially relevant to the VCM side effect of haloperidol treatment. In addition, our previous studies had prioritized several other glutamate receptors such as \textit{Grin1} and \textit{Grin2a} as candidates that response to haloperidol treatment using independent data and methods \citep{crowley2012antipsychotic}. \\

Similar studies were conducted for the two F1 mice of 129S1Sv/ImJ$\times$PWK/PhJ. To better map RNA-seq reads, we first built two pseudogenomes for 129S1Sv/ImJ and PWK/PhJ by incorporating Sanger SNPs and indels \citep{keane2011mouse} into reference genome and mapped the RNA-seq reads to the two pseudogenomes separately. Then alignments were remapped back to the reference coordinate system and the observed genetic variants were annotated for each RNA-seq fragment. IsoDOT identified much less DIU or DIE genes in these two F1 mice than in the two C57BL/6J mice. At FDR 0.2, no DIU genes were identified and 85 DIE gene were identified. Six of these 85 genes were involved in actin-binding, though this functional category was not significantly overrepresented. \\

The greater level of DIE and DIU in C57BL/6J than 129S1Sv/ImJ$\times$PWK/PhJ following chronic haloperidol treatment was consistent with behavioral phenotype data which showed that C57BL6/J mice had greater susceptibility to haloperidol-induced VCM (Supplementary Figure 19) \citep{crowley2012antipsychotic}. Therefore, some of the C57BL/6J transcriptional changes detailed in this study might contribute to the development of haloperidol-induced VCM.

\subsection*{Allele-specific differential isoform usage.} About 37.2\% of the RNA-seq fragments from the two 129S1Sv/ImJ$\times$PWK/PhJ F1 mice were only mapped to paternal or maternal allele or were mapped to one allele with less mismatch, and hence they were allele-specific RNA-seq fragments. There was no genome-wide bias, i.e., $\sim$50\% of the allele-specific RNA-seq fragments were mapped to each parental strain (Supplementary Table 1). Using these allele-specific RNA-seq fragments, we applied IsoDOT to assess differential isoform usage between maternal and paternal alleles. At a liberal FDR cutoff 0.2, no DIU gene was identified, and 19 or 30 DIE genes were identified from the haloperidol/placebo treated F1 mice, respectively. The genes with significant differential isoform usage in the haloperidol treated mice, but not the placebo treated mice might indicate genetic$\times$treatment interactions. Supplementary Table 3 listed 23 such genes with DIU p-values $<$ 0.01 in the haloperidol treated mice, and DIU p-values $>$ 0.1 in the placebo treated mice. Among them, \textit{Synpo} and \textit{Snap25} are associated with neuron functions. \textit{Snap25} is known to be associated with schizophrenia and/or haloperidol treatment at DNA \citep{muller2005snap}, RNA \citep{sommer2010differential}, and protein levels \citep{thompson1998altered,honer2002abnormalities}. To the best of our knowledge, this is the first report that the differential isoform usage of \textit{Snap25} is associated with genetic$\times$haloperidol treatment interaction. 

\subsection*{Software and data availability.}
An R package of IsoDOT is available at \url{http://www.bios.unc.edu/~weisun/software/isoform.htm}. Testing differential isoform expression/usage is computationally intensive. Using IsoDOT with up to 1,000 parametric bootstrap, it will take on average 1-3 minutes to test differential isoform usage for a gene on a single processor. Parallel computation is needed and straightforward for genome-wide study. We are also actively working on implementing our method using Graphics Processing Unit (GPU) using the massively parallel algorithm described in Supplementary Materials Section E. Simulated RNA-seq data can be downloaded from \url{http://www.bios.unc.edu/~weisun/software/isoform_files/}. The RNA-seq data of mouse haloperidol treatment experiment are available from NCBI GEO (GSExxx). 

\subsection*{Discussion}

We have developed a new statistical method named IsoDOT to assess differential isoform expression or usage from RNA-seq data with respect to categorical or continuous covariates. {The resampling based approach is the basis of our hypothesis testing method. Two components of our method, the negative binomial distribution assumption and the Log penalty for penalized estimation are important for the success of this resampling based approach. First, negative binomial distribution is a well-accepted choice to model RNA-seq fragment data across biological replicates. For the self-completeness of our paper, we also demonstrate that negative binomial distribution can provide good fit of RNA-seq fragment count data, while Poisson distribution assumption leads to severe underestimate of variance (Supplementary Figure 20). Replacing the Log penalty with the Lasso penalty in IsoDOT leads to inaccurate type I error control and/or reduced power (Supplementary Figure 21). The limitation of Lasso is especially apparent when we do not use isoform annotation, where the number of candidate isoforms is much larger than sample size. This is consistent with the previous findings that the Lasso tends to select more false positives and has larger bias in effect estimates than the Log penalty \citep{Chen2013}. } \\

Some biases should be accounted to obtain better estimates of RNA isoform abundance. For example, the RNA-seq reads may not be uniformly distributed along the transcript, and DNA sequence features such as GC content may affect the abundance RNA-seq reads. Such biases, if exits, affect both the likelihoods under null and alternative and do not alter our Type I error rate; furthermore, it may have limited impact to the power of testing  because we calculate p-values by resampling approaches. For example, if the last exon of a gene tends to have larger number of RNA-seq reads, our method may overestimate the expression of the isoform harboring the last exon. However, such over-estimation occurs under both null and alternative and does not lead to inflated or deflated type I error. Systematically accounting for such bias is among our future works to further improve the performance of IsoDOT. \\

An important contribution of IsoDOT is to allow assessing differential isoform usage between one case and one control sample. This is especially important for paired samples, e.g., maternal and paternal alleles of one individual. When there are multiple paired samples, we can combine the p-values of multiple pairs using meta-analysis via Fisher's method or Stouffer's Z-score \citep{hunter1982meta}. In the preliminary study reported in this paper, we have identified several interesting genes (\textit{Utrn}, \textit{Dmd}, \textit{Grin2b}, \textit{Snap25}) whose isoform usage may respond to haloperidol treatment or its side effect. In the near future, we plan to extend this study to include larger number of mice with diverse genetic backgrounds including mice from the Collaborative Cross \citep{churchill2004collaborative,consortium2012genome}. \\

Recently developed sequencing techniques can deliver longer reads such as 2 x 250 bp reads from Illumina's MiSeq or 400 bp reads from Ion Torrent. Our methods can handle these longer sequencing reads without any difficulty. When the RNA-seq reads are long enough, transcriptome reconstruction becomes easier. However, until all the isoforms of a transcript cluster can be unambiguously reconstructed and all the RNA-seq fragments can be assigned to an isoform with (almost) 100\% certainty, testing differential isoform expression/usage remains a challenging problem where methods and software such as IsoDOT are needed.  Another recent development of RNA-seq technique is to deliver ``stranded'' sequences so that RNA-seq from sense and antisense strands can be separated. For the analysis of such stranded RNA-seq data, the only step of IsoDOT pipeline that needs to be modified is to count the RNA-seq fragments for sense and anti-sense strands separately. \\

This work was partially supported by NIH grants GM105785, CA167684, CA149569, MH101819, GM074175 and P50 HG006582.

\clearpage

\bibliography{asSeq} 

\begin{thebibliography}{54}
\providecommand{\natexlab}[1]{#1}
\providecommand{\url}[1]{\texttt{#1}}
\expandafter\ifx\csname urlstyle\endcsname\relax
  \providecommand{\doi}[1]{doi: #1}\else
  \providecommand{\doi}{doi: \begingroup \urlstyle{rm}\Url}\fi

\bibitem[Alamancos et~al.(2014)Alamancos, Agirre, and
  Eyras]{alamancos2014methods}
Gael~P Alamancos, Eneritz Agirre, and Eduardo Eyras.
\newblock Methods to study splicing from high-throughput {RNA} sequencing data.
\newblock In \emph{Spliceosomal Pre-mRNA Splicing}, pages 357--397. Springer,
  2014.

\bibitem[Anders and Huber(2010)]{anders2010differential}
Simon Anders and Wolfgang Huber.
\newblock Differential expression analysis for sequence count data.
\newblock \emph{Genome Biol}, 11\penalty0 (10):\penalty0 R106, 2010.

\bibitem[Anders et~al.(2012)Anders, Reyes, and Huber]{anders2012detecting}
Simon Anders, Alejandro Reyes, and Wolfgang Huber.
\newblock Detecting differential usage of exons from {RNA}-seq data.
\newblock \emph{Genome research}, 22\penalty0 (10):\penalty0 2008--2017, 2012.

\bibitem[Barbosa-Morais et~al.(2012)Barbosa-Morais, Irimia, Pan, Xiong,
  Gueroussov, Lee, Slobodeniuc, Kutter, Watt, {\c{C}}olak,
  et~al.]{barbosa2012evolutionary}
N.L. Barbosa-Morais, M.~Irimia, Q.~Pan, H.Y. Xiong, S.~Gueroussov, L.J. Lee,
  V.~Slobodeniuc, C.~Kutter, S.~Watt, R.~{\c{C}}olak, et~al.
\newblock The evolutionary landscape of alternative splicing in vertebrate
  species.
\newblock \emph{Science}, 338\penalty0 (6114):\penalty0 1587--1593, 2012.

\bibitem[Bohnert and R{\"a}tsch(2010)]{bohnert2010rquant}
R.~Bohnert and G.~R{\"a}tsch.
\newblock rquant. web: a tool for {RNA}-seq-based transcript quantitation.
\newblock \emph{Nucleic acids research}, 38\penalty0 (suppl 2):\penalty0
  W348--W351, 2010.

\bibitem[Bullard et~al.(2010)Bullard, Purdom, Hansen, and
  Dudoit]{bullard2010evaluation}
James~H Bullard, Elizabeth Purdom, Kasper~D Hansen, and Sandrine Dudoit.
\newblock Evaluation of statistical methods for normalization and differential
  expression in mrna-seq experiments.
\newblock \emph{BMC bioinformatics}, 11\penalty0 (1):\penalty0 94, 2010.

\bibitem[Chen(2012)]{chen2012statistical}
L.~Chen.
\newblock Statistical and computational methods for high-throughput sequencing
  data analysis of alternative splicing.
\newblock \emph{Statistics in Biosciences}, pages 1--18, 2012.

\bibitem[Chen et~al.(2013)Chen, Sun, and J.]{Chen2013}
T.H. Chen, W.~Sun, and Fine J.
\newblock Designing penalty functions in high dimensional problems: The role of
  tuning parameters.
\newblock UNC Chapel Hill, 2013.

\bibitem[Churchill et~al.(2004)Churchill, Airey, Allayee, Angel, Attie, Beatty,
  Beavis, Belknap, Bennett, Berrettini, et~al.]{churchill2004collaborative}
G.A. Churchill, D.C. Airey, H.~Allayee, J.M. Angel, A.D. Attie, J.~Beatty, W.D.
  Beavis, J.K. Belknap, B.~Bennett, W.~Berrettini, et~al.
\newblock The collaborative cross, a community resource for the genetic
  analysis of complex traits.
\newblock \emph{Nature genetics}, 36\penalty0 (11):\penalty0 1133--1137, 2004.

\bibitem[Consortium(2012)]{consortium2012genome}
C.C. Consortium.
\newblock The genome architecture of the collaborative cross mouse genetic
  reference population.
\newblock \emph{Genetics}, 190:\penalty0 389--401, 2012.

\bibitem[Crowley et~al.(2012)Crowley, Adkins, Pratt, Quackenbush, van~den Oord,
  Moy, Wilhelmsen, Cooper, Bogue, McLeod, et~al.]{crowley2012antipsychotic}
JJ~Crowley, DE~Adkins, AL~Pratt, CR~Quackenbush, EJ~van~den Oord, SS~Moy,
  KC~Wilhelmsen, TB~Cooper, MA~Bogue, HL~McLeod, et~al.
\newblock Antipsychotic-induced vacuous chewing movements and extrapyramidal
  side effects are highly heritable in mice.
\newblock \emph{The pharmacogenomics journal}, 12\penalty0 (2):\penalty0 147,
  2012.

\bibitem[Denoeud et~al.(2008)Denoeud, Aury, Da~Silva, Noel, Rogier, Delledonne,
  Morgante, Valle, Wincker, Scarpelli, et~al.]{denoeud2008annotating}
F.~Denoeud, J.M. Aury, C.~Da~Silva, B.~Noel, O.~Rogier, M.~Delledonne,
  M.~Morgante, G.~Valle, P.~Wincker, C.~Scarpelli, et~al.
\newblock Annotating genomes with massive-scale {RNA} sequencing.
\newblock \emph{Genome Biol}, 9\penalty0 (12):\penalty0 R175, 2008.

\bibitem[Djebali et~al.(2012)Djebali, Davis, Merkel, Dobin, Lassmann,
  Mortazavi, Tanzer, Lagarde, Lin, Schlesinger, et~al.]{djebali2012landscape}
S.~Djebali, C.A. Davis, A.~Merkel, A.~Dobin, T.~Lassmann, A.~Mortazavi,
  A.~Tanzer, J.~Lagarde, W.~Lin, F.~Schlesinger, et~al.
\newblock Landscape of transcription in human cells.
\newblock \emph{Nature}, 489\penalty0 (7414):\penalty0 101--108, 2012.

\bibitem[Flicek et~al.(2011)Flicek, Amode, Barrell, Beal, Brent, Chen, Clapham,
  Coates, Fairley, Fitzgerald, et~al.]{Flicek11}
P.~Flicek, M.R. Amode, D.~Barrell, K.~Beal, S.~Brent, Y.~Chen, P.~Clapham,
  G.~Coates, S.~Fairley, S.~Fitzgerald, et~al.
\newblock {Ensembl 2011}.
\newblock \emph{Nucleic acids research}, 39\penalty0 (suppl 1):\penalty0 D800,
  2011.
\newblock ISSN 0305-1048.

\bibitem[Glaus et~al.(2012)Glaus, Honkela, and Rattray]{glaus2012identifying}
Peter Glaus, Antti Honkela, and Magnus Rattray.
\newblock Identifying differentially expressed transcripts from {RNA}-seq data
  with biological variation.
\newblock \emph{Bioinformatics}, 28\penalty0 (13):\penalty0 1721--1728, 2012.

\bibitem[Honer et~al.(2002)Honer, Falkai, Bayer, Xie, Hu, Li, Arango, Mann,
  Dwork, and Trimble]{honer2002abnormalities}
W.G. Honer, P.~Falkai, T.A. Bayer, J.~Xie, L.~Hu, H.Y. Li, V.~Arango, J.J.
  Mann, A.J. Dwork, and W.S. Trimble.
\newblock Abnormalities of snare mechanism proteins in anterior frontal cortex
  in severe mental illness.
\newblock \emph{Cerebral Cortex}, 12\penalty0 (4):\penalty0 349--356, 2002.

\bibitem[Hunter et~al.(1982)Hunter, Schmidt, and Jackson]{hunter1982meta}
J.E. Hunter, F.L. Schmidt, and G.B. Jackson.
\newblock \emph{Meta-analysis}.
\newblock Sage Publ., 1982.

\bibitem[Jiang and Salzman(2012)]{jiang2012statistical}
H.~Jiang and J.~Salzman.
\newblock Statistical properties of an early stopping rule for resampling-based
  multiple testing.
\newblock \emph{Biometrika}, 99\penalty0 (4):\penalty0 973--980, 2012.

\bibitem[Jiang and Wong(2009)]{Jiang09}
H.~Jiang and W.H. Wong.
\newblock {Statistical inferences for isoform expression in {RNA}-Seq}.
\newblock \emph{Bioinformatics}, 25\penalty0 (8):\penalty0 1026, 2009.
\newblock ISSN 1367-4803.

\bibitem[Katz et~al.(2010)Katz, Wang, Airoldi, and Burge]{katz2010analysis}
Y.~Katz, E.T. Wang, E.M. Airoldi, and C.B. Burge.
\newblock Analysis and design of {RNA} sequencing experiments for identifying
  isoform regulation.
\newblock \emph{Nature methods}, 7\penalty0 (12):\penalty0 1009--1015, 2010.

\bibitem[Keane et~al.(2011)Keane, Goodstadt, Danecek, White, Wong, Yalcin,
  Heger, Agam, Slater, Goodson, et~al.]{keane2011mouse}
T.M. Keane, L.~Goodstadt, P.~Danecek, M.A. White, K.~Wong, B.~Yalcin, A.~Heger,
  A.~Agam, G.~Slater, M.~Goodson, et~al.
\newblock Mouse genomic variation and its effect on phenotypes and gene
  regulation.
\newblock \emph{Nature}, 477\penalty0 (7364):\penalty0 289--294, 2011.

\bibitem[Langmead et~al.(2010)Langmead, Hansen, and Leek]{Langmead10}
B.~Langmead, K.D. Hansen, and J.T. Leek.
\newblock {Cloud-scale {RNA}-sequencing differential expression analysis with
  Myrna}.
\newblock \emph{Genome biology}, 11\penalty0 (8):\penalty0 R83, 2010.

\bibitem[Leng et~al.(2013)Leng, Dawson, Thomson, Ruotti, Rissman, Smits, Haag,
  Gould, Stewart, and Kendziorski]{leng2013ebseq}
Ning Leng, John~A Dawson, James~A Thomson, Victor Ruotti, Anna~I Rissman,
  Bart~MG Smits, Jill~D Haag, Michael~N Gould, Ron~M Stewart, and Christina
  Kendziorski.
\newblock {EBSeq}: an empirical bayes hierarchical model for inference in
  {RNA}-seq experiments.
\newblock \emph{Bioinformatics}, 29\penalty0 (8):\penalty0 1035--1043, 2013.

\bibitem[Li et~al.(2010)Li, Ruotti, Stewart, Thomson, and Dewey]{li2010rna}
B.~Li, V.~Ruotti, R.M. Stewart, J.A. Thomson, and C.N. Dewey.
\newblock {RNA}-seq gene expression estimation with read mapping uncertainty.
\newblock \emph{Bioinformatics}, 26\penalty0 (4):\penalty0 493--500, 2010.

\bibitem[Li et~al.(2011{\natexlab{a}})Li, Jiang, Hu, Brown, Huang, and
  Bickel]{li2011sparse}
J.J. Li, C.R. Jiang, Y.~Hu, B.J. Brown, H.~Huang, and P.J. Bickel.
\newblock Sparse linear modeling of {RNA}-seq data for isoform discovery and
  abundance estimation.
\newblock \emph{Proc Natl Acad Sci. USA}, in press, 2011{\natexlab{a}}.

\bibitem[Li et~al.(2011{\natexlab{b}})Li, Feng, and Jiang]{li2011isolasso}
W.~Li, J.~Feng, and T.~Jiang.
\newblock {Isolasso: a lasso regression approach to {RNA}-seq based
  transcriptome assembly}.
\newblock \emph{Research in Computational Molecular Biology}, pages 168--188,
  2011{\natexlab{b}}.

\bibitem[Mazumder et~al.(2011)Mazumder, Friedman, and
  Hastie]{mazumder2011sparsenet}
Rahul Mazumder, Jerome~H Friedman, and Trevor Hastie.
\newblock Sparsenet: Coordinate descent with nonconvex penalties.
\newblock \emph{Journal of the American Statistical Association}, 106\penalty0
  (495):\penalty0 1125--1138, 2011.

\bibitem[Mortazavi et~al.(2008)Mortazavi, Williams, McCue, Schaeffer, and
  Wold]{mortazavi2008mapping}
Ali Mortazavi, Brian~A Williams, Kenneth McCue, Lorian Schaeffer, and Barbara
  Wold.
\newblock Mapping and quantifying mammalian transcriptomes by {RNA}-seq.
\newblock \emph{Nature methods}, 5\penalty0 (7):\penalty0 621--628, 2008.

\bibitem[M{\"u}ller et~al.(2005)M{\"u}ller, Klempan, De~Luca, Sicard, Volavka,
  Czobor, Sheitman, Lindenmayer, Citrome, McEvoy, et~al.]{muller2005snap}
D.J. M{\"u}ller, T.A. Klempan, V.~De~Luca, T.~Sicard, J.~Volavka, P.~Czobor,
  B.B. Sheitman, J.P. Lindenmayer, L.~Citrome, J.P. McEvoy, et~al.
\newblock The snap-25 gene may be associated with clinical response and weight
  gain in antipsychotic treatment of schizophrenia.
\newblock \emph{Neuroscience letters}, 379\penalty0 (2):\penalty0 81, 2005.

\bibitem[Pachter(2011)]{pachter2011models}
L.~Pachter.
\newblock Models for transcript quantification from {RNA}-seq.
\newblock \emph{Arxiv preprint arXiv:1104.3889}, 2011.

\bibitem[Pan et~al.(2008)Pan, Shai, Lee, Frey, and Blencowe]{Pan08}
Q.~Pan, O.~Shai, L.J. Lee, B.J. Frey, and B.J. Blencowe.
\newblock {Deep surveying of alternative splicing complexity in the human
  transcriptome by high-throughput sequencing}.
\newblock \emph{Nature genetics}, 40\penalty0 (12):\penalty0 1413--1415, 2008.
\newblock ISSN 1061-4036.

\bibitem[Purdom et~al.(2008)Purdom, Simpson, Robinson, Conboy, Lapuk, and
  Speed]{purdom2008firma}
E~Purdom, Ken~M Simpson, Mark~D Robinson, JG~Conboy, AV~Lapuk, and Terence~P
  Speed.
\newblock Firma: a method for detection of alternative splicing from exon array
  data.
\newblock \emph{Bioinformatics}, 24\penalty0 (15):\penalty0 1707--1714, 2008.

\bibitem[Rashid et~al.(2011)Rashid, Giresi, Ibrahim, Sun, and
  Lieb]{rashid2011zinba}
Naim~U Rashid, Paul~G Giresi, Joseph~G Ibrahim, Wei Sun, and Jason~D Lieb.
\newblock Zinba integrates local covariates with dna-seq data to identify broad
  and narrow regions of enrichment, even within amplified genomic regions.
\newblock \emph{Genome Biology}, 12\penalty0 (7):\penalty0 R67, 2011.

\bibitem[Richard et~al.(2010)Richard, Schulz, Sultan, N{\"u}rnberger,
  Schrinner, Balzereit, Dagand, Rasche, Lehrach, Vingron,
  et~al.]{richard2010prediction}
H.~Richard, M.H. Schulz, M.~Sultan, A.~N{\"u}rnberger, S.~Schrinner,
  D.~Balzereit, E.~Dagand, A.~Rasche, H.~Lehrach, M.~Vingron, et~al.
\newblock Prediction of alternative isoforms from exon expression levels in
  {RNA}-seq experiments.
\newblock \emph{Nucleic Acids Research}, 38\penalty0 (10):\penalty0 e112--e112,
  2010.

\bibitem[Roberts et~al.(2011)Roberts, Trapnell, Donaghey, Rinn, Pachter,
  et~al.]{roberts2011improving}
A.~Roberts, C.~Trapnell, J.~Donaghey, J.L. Rinn, L.~Pachter, et~al.
\newblock Improving {RNA}-seq expression estimates by correcting for fragment
  bias.
\newblock \emph{Genome biology}, 12\penalty0 (3):\penalty0 R22, 2011.

\bibitem[Robinson et~al.(2010)Robinson, McCarthy, and Smyth]{robinson2010edger}
Mark~D Robinson, Davis~J McCarthy, and Gordon~K Smyth.
\newblock {edgeR}: a {Bioconductor} package for differential expression
  analysis of digital gene expression data.
\newblock \emph{Bioinformatics}, 26\penalty0 (1):\penalty0 139--140, 2010.

\bibitem[Salzman et~al.(2011)Salzman, Jiang, and Wong]{salzman2011statistical}
J.~Salzman, H.~Jiang, and W.H. Wong.
\newblock Statistical modeling of {RNA}-seq data.
\newblock \emph{Statistical Science}, 26\penalty0 (1):\penalty0 62--83, 2011.

\bibitem[Sherman et~al.(2009)Sherman, Lempicki, et~al.]{sherman2009systematic}
BT~Sherman, RA~Lempicki, et~al.
\newblock Systematic and integrative analysis of large gene lists using david
  bioinformatics resources.
\newblock \emph{Nature protocols}, 4\penalty0 (1):\penalty0 44, 2009.

\bibitem[Sommer et~al.(2010)Sommer, Schmitt, Heck, Schaeffer, Fendt, Zink,
  Nieselt, Symons, Petroianu, Lex, et~al.]{sommer2010differential}
J.U. Sommer, A.~Schmitt, M.~Heck, EL~Schaeffer, M.~Fendt, M.~Zink, K.~Nieselt,
  S.~Symons, G.~Petroianu, A.~Lex, et~al.
\newblock Differential expression of presynaptic genes in a rat model of
  postnatal hypoxia: relevance to schizophrenia.
\newblock \emph{European archives of psychiatry and clinical neuroscience},
  260:\penalty0 81--89, 2010.

\bibitem[Sun et~al.(2010)Sun, Ibrahim, and Zou]{Sun10}
W.~Sun, J.G. Ibrahim, and F.~Zou.
\newblock {Genomewide Multiple-Loci Mapping in Experimental Crosses by
  Iterative Adaptive Penalized Regression}.
\newblock \emph{Genetics}, 185\penalty0 (1):\penalty0 349, 2010.

\bibitem[Sun(2012)]{sun2012statistical}
Wei Sun.
\newblock A statistical framework for eqtl mapping using {RNA}-seq data.
\newblock \emph{Biometrics}, 68\penalty0 (1):\penalty0 1--11, 2012.

\bibitem[Thompson et~al.(1998)Thompson, Sower, and
  Perrone-Bizzozero]{thompson1998altered}
P.M. Thompson, A.C. Sower, and N.I. Perrone-Bizzozero.
\newblock Altered levels of the synaptosomal associated protein snap-25 in
  schizophrenia.
\newblock \emph{Biological psychiatry}, 43\penalty0 (4):\penalty0 239--243,
  1998.

\bibitem[Thorisson et~al.(2005)Thorisson, Smith, Krishnan, and
  Stein]{Thorisson2005}
G.A. Thorisson, A.V. Smith, L.~Krishnan, and L.D. Stein.
\newblock {The international HapMap project web site}.
\newblock \emph{Genome research}, 15\penalty0 (11):\penalty0 1592, 2005.

\bibitem[Tibshirani(1996)]{Tibshirani96}
R.~Tibshirani.
\newblock {Regression shrinkage and selection via the lasso}.
\newblock \emph{Journal of the Royal Statistical Society. Series B
  (Methodological)}, 58\penalty0 (1):\penalty0 267--288, 1996.

\bibitem[Trapnell et~al.(2009)Trapnell, Pachter, and Salzberg]{Trapnell09}
C.~Trapnell, L.~Pachter, and S.L. Salzberg.
\newblock {TopHat: discovering splice junctions with {RNA}-Seq}.
\newblock \emph{Bioinformatics}, 25\penalty0 (9):\penalty0 1105, 2009.

\bibitem[Trapnell et~al.(2010)Trapnell, Williams, Pertea, Mortazavi, Kwan, van
  Baren, Salzberg, Wold, and Pachter]{Trapnell10}
C.~Trapnell, B.A. Williams, G.~Pertea, A.~Mortazavi, G.~Kwan, M.J. van Baren,
  S.L. Salzberg, B.J. Wold, and L.~Pachter.
\newblock {Transcript assembly and quantification by {RNA}-Seq reveals
  unannotated transcripts and isoform switching during cell differentiation}.
\newblock \emph{Nature biotechnology}, 28\penalty0 (5):\penalty0 511--515,
  2010.

\bibitem[Trapnell et~al.(2013)Trapnell, Hendrickson, Sauvageau, Goff, Rinn, and
  Pachter]{trapnell2013differential}
Cole Trapnell, David~G Hendrickson, Martin Sauvageau, Loyal Goff, John~L Rinn,
  and Lior Pachter.
\newblock Differential analysis of gene regulation at transcript resolution
  with {RNA}-seq.
\newblock \emph{Nature biotechnology}, 31\penalty0 (1):\penalty0 46--53, 2013.

\bibitem[Wang et~al.(2008)Wang, Sandberg, Luo, Khrebtukova, Zhang, Mayr,
  Kingsmore, Schroth, and Burge]{Wang08Alt}
E.T. Wang, R.~Sandberg, S.~Luo, I.~Khrebtukova, L.~Zhang, C.~Mayr, S.F.
  Kingsmore, G.P. Schroth, and C.B. Burge.
\newblock {Alternative isoform regulation in human tissue transcriptomes}.
\newblock \emph{Nature}, 456\penalty0 (7221):\penalty0 470--476, 2008.
\newblock ISSN 0028-0836.

\bibitem[Wang and Cooper(2007)]{Wang07splicing}
G.S. Wang and T.A. Cooper.
\newblock Splicing in disease: disruption of the splicing code and the decoding
  machinery.
\newblock \emph{Nature Reviews Genetics}, 8\penalty0 (10):\penalty0 749--761,
  2007.

\bibitem[Wang et~al.(2009)Wang, Gerstein, and Snyder]{Wang2009rna}
Z.~Wang, M.~Gerstein, and M.~Snyder.
\newblock {RNA-Seq: a revolutionary tool for transcriptomics}.
\newblock \emph{Nature Reviews Genetics}, 10\penalty0 (1):\penalty0 57--63,
  2009.
\newblock ISSN 1471-0056.

\bibitem[Xia et~al.(2011)Xia, Wen, Chang, and Zhou]{xia2011nsmap}
Z.~Xia, J.~Wen, C.C. Chang, and X.~Zhou.
\newblock Nsmap: A method for spliced isoforms identification and
  quantification from {RNA}-seq.
\newblock \emph{BMC bioinformatics}, 12\penalty0 (1):\penalty0 162, 2011.

\bibitem[Xing et~al.(2006)Xing, Yu, Wu, Roy, Kim, and Lee]{xing2006expectation}
Y.~Xing, T.~Yu, Y.N. Wu, M.~Roy, J.~Kim, and C.~Lee.
\newblock An expectation-maximization algorithm for probabilistic
  reconstructions of full-length isoforms from splice graphs.
\newblock \emph{Nucleic acids research}, 34\penalty0 (10):\penalty0 3150, 2006.

\bibitem[Zhao and Yu(2006)]{zhao2006model}
P.~Zhao and B.~Yu.
\newblock On model selection consistency of lasso.
\newblock \emph{The Journal of Machine Learning Research}, 7:\penalty0
  2541--2563, 2006.

\bibitem[Zou(2006)]{zou2006adaptive}
H.~Zou.
\newblock The adaptive lasso and its oracle properties.
\newblock \emph{Journal of the American Statistical Association}, 101\penalty0
  (476):\penalty0 1418--1429, 2006.

\end{thebibliography}

\clearpage

\begin{center}
\textbf{\Large{Supplementary Materials}}
\end{center}

\appendix
\setcounter{figure}{0} \renewcommand{\thefigure}{S\arabic{figure}}
\setcounter{table}{0} \renewcommand{\thetable}{S\arabic{table}}

% ========================================================
\section{Calculation of effective length}
% ========================================================

An RNA-seq fragment is a segment of RNA to be sequenced. Usually only part of an RNA-seq fragment is sequenced: one end or both ends, hence single-end sequencing or paired-end sequencing. All the discussions in this section are for paired-end reads, though the extension to single-end reads is straightforward. The minimum fragment size is the read length, denoted by $d$. This happens when the two reads of a fragment completely overlap. We impose an upper bound for the fragment length based on prior knowledge of the experimental procedure and denote the upper bound by $l_M$. Then the fragment length $l$ satisfies $d \leq l \leq l_M$. We denote the distribution of the fragment length for sample $i$ by $\varphi_i(l)$, which can be calculated using observed read alignment information. \\

\begin{figure}[ht]
  \centering
  \includegraphics[scale=0.5]{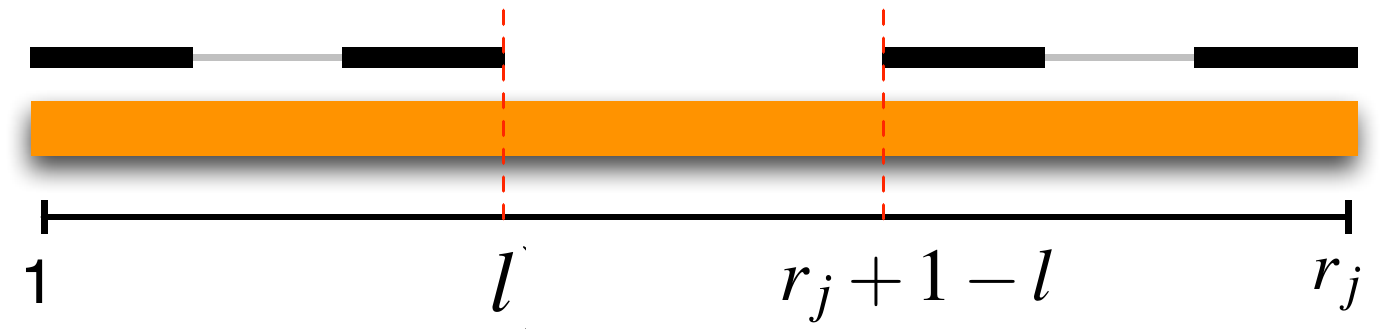}
  \caption{An illustration of effective length calculation for an exon of $r_j$ bps and RNA-seq fragment of $l$ bps. The orange box indicates the exon, and the black lines above the orange box indicate two RNA-seq fragments, while each RNA-seq fragment is sequenced by a paired-end read. There are $r_j +1 -l$ distinct choices to select an RNA-seq fragment of $l$ bps from this exon, and thus the effective length is $r_j +1 -l$.}
   \label{fig7:effLen}
\end{figure}

For the $i$-th sample, the effective length of exon $j$ of $r_j$ base pairs (bps) is
\begin{eqnarray*}
 \eta_{i,\{j\}} = f(r_j, d, l_M, \varphi_i) =
   \begin{cases}
    0 & \quad \text{if } r_j < d \\
    \displaystyle \sum_{l=d}^{\min(r_j, l_M)} \varphi_i(l) (r_j +1 - l) &
    \quad \text{if $r_j \geq d$}
  \end{cases}.
\end{eqnarray*}
If $r_j < d$, the exon is shorter than the shortest fragment length, and thus the effective length of this exon is 0. In other words, no RNA-seq fragment is expected to overlap and only overlap with this exon. If $r_j \geq d$, the effective length is $r_j +1 - l$, i.e., there are $r_j +1 - l$ distinct RNA-seq fragments that can be sequenced from this exon (Figure~\ref{fig7:effLen}). Then $\sum_{l=d}^{\min(r_j, l_M)} \varphi_i(l) (r_j +1 - l)$ is summation across all likely fragment lengths, weighted by the probability of having fragment length $l$.\\

In the following discussions, to simplify the notation, we skip the subscript of $i$. For two exons $j$ and $k$ ($j < k$) of lengths $r_j$ and $r_k$, which are adjacent in the transcript, the effective length for the fragments that cover both exons is
\begin{eqnarray}
 \eta_{\{j,k\}} = f(r_j+r_k, d, l_M, \varphi) - \eta_{\{j\}} - \eta_{\{k\}}.
\end{eqnarray}
For three exons $j$, $h$, and $k$ ($j < h < k$) of lengths $r_j$, $r_h$ and $r_k$, which are adjacent in the transcript, the effective length for the fragments that cover all three exons is
\begin{eqnarray*}
  \eta_{\{j, h, k\}} &=& f(r_j+r_h+r_k, d, l_M) - \eta_{\{j,h\}}  - \eta_{\{h,k\}}  - \eta_{\{j,(h),k\}} - \eta_{\{j\}} - \eta_{\{h\}} - \eta_{\{k\}},
\end{eqnarray*}
where $\eta_{\{j, (h), k\}}$ is the effective length in the scenario that the transcript covers consecutive exons $j$, $h$, and $k$, whereas the observed paired-end read only covers exons $j$ and $k$.
\begin{eqnarray*}
  \eta_{\{j, (h), k\}} = \begin{cases}
    0 & \quad \text{if } (r_j, r_h, r_k) \in R_1  \\
    \displaystyle \sum_{l=2d + r_h}^{\min(r_j+r_h+r_k, l_M)} \varphi(l) \delta_l &
    \quad \text{otherwise}
  \end{cases}
\end{eqnarray*}
where $R_1 = \{ (r_j, r_h, r_k):\  r_j < d \textrm{ or } r_k < d \textrm{ or } r_h + 2d > l_M\}$, and $\delta_l = \min(r_j, l - r_h - d) - \max(d, l - r_h - r_k) + 1$. The above formula is derived by the following arguments. Let $l_j$ and $l_k$ be the lengths of the parts of the fragment that overlaps with exon $j$ and $k$, respectively. Given $l$, the restriction of $l_j$ and $l_k$ are $l = l_j + l_k + r_h$, $d \leq l_j \leq r_j$, and $d \leq l_k \leq r_k$, and thus the range of $l_j$ is $\max(d, l - r_h - r_k) \leq l_j \leq \min(r_j, l - r_h -d)$. For more than 3 consecutive exons, the effective lengths can be calculated using recursive calls to the above equations.\\

In practice, a few sequence fragments may be observed even when the effective length is zero, which may be due to sequencing errors. To improve the robustness of our method, we modify the design matrix $\X$ by adding a pre-determined constant $\texttt{eLenMin}$ to each element of $\X$. \\

% ========================================================
\section{Selection of candidate isoforms}
% ========================================================

For each gene, we select a set of candidate isoforms given the fragment counts at each exon set. We define a start exon as an exon that is only connected to downstream exons and an end exon as an exon that is only connected to upstream exons. An initial set of start and end exons can be identified simply by examining the observed exon sets.\\

Next, we seek to find more start and end exons by identifying break points where the read-depth of two adjacent exons are different. Specifically, suppose the gene of interest has $h$ exons. Let $y_{\{k\}}$ be the number of fragments overlapping the $k$th exon of this gene.  We apply a Pearson chi-squared test to assess whether the frequency distribution of $y_{\{k-1\}}$ and $y_{\{k\}}$ is significantly different from theoretical expectation based on their effective lengths. For $k=2$, ..., $h$, there are $h-1$ possible break points, which correspond to $h-1$ p-values: $\texttt{pB}_1$, ..., $\texttt{pB}_h$. We order those possible break points by the corresponding p-values in ascending order and select the top $$\min\left(\texttt{maxBreaks}, \ \sum_{k=2}^h I (\texttt{pB}_k < \texttt{pvalBreaks})\right)$$ break points, where $I(\cdot)$ is an indicator function, \texttt{maxBreaks} and \texttt{pvalBreaks} are two pre-set parameters. \texttt{maxBreaks} is the maximum number of break points, with default value 5, and \texttt{pvalBreaks} is a p-value cutoff, with default value 0.05.  If the $k$-th break points is selected, the $(k-1)$th exon is added to the set of start exons and the $k$th exon is added to the set of end exons. After identifying all possible start and end exons, we can construct all isoforms that have consecutive exons. \\

For each exon set, we assign a p-value to quantify whether it is expressed. Suppose there are $n_T$ fragments for the gene of interest and among them $n_j$ fragments are from the $j$th exon set. Then the expression p-value is $\texttt{pE}_j = \texttt{pbinom}(n_j, n_T, l_j/l_T)$,  where $\texttt{pbinom}(\cdot, n, \pi)$ is cumulative binomial distribution function with $n$ trials and probability of success $\pi$, $l_j$ is the effective length of the $j$th exon set, and $l_T$ is the total effective length of this gene. We claim the $j$th exon set is expressed if
  $$\texttt{pE}_j  \geq \texttt{pvalExpress} \textrm{ and } \frac{n_j/l_j}{n_T/l_T} > \texttt{foldExpress},$$
 where \texttt{pvalExpress} and \texttt{foldExpress} are two pre-set parameters, with default values 0.01 and 1/5, respectively. \\
 
 Finally, we select all the expressed exon sets that harbor at least on exon-skipping event, and order them by the $\texttt{pE}_j $ in a descending order. Then for each of these ordered exon sets, we construct new RNA-isoforms by inserting this exon set into each existing isoform if this exon set is compatible with the isoform. We stop adding more isoforms if either 
 $$q/m > \texttt{pMaxRel} \textrm{ or } q > \texttt{pMaxAbs}$$
 where $q$ is the number of isoforms, \texttt{pMaxRel} and \texttt{pMaxAbs} are pre-set parameters with default values 10 and 2000, respectively. In oder words, we allow the number of isoforms to be at most 10 times the number of exon sets and the total number of isoforms to be at most 2000. Users can change these default values. Our penalized regression can handle the situation \texttt{pMaxRel}=100 and \texttt{pMaxAbs}=100,000; however it may significantly reduce the computational efficiency. \\
 
% ========================================================
\section{Model fitting of the penalized negative binomial regression}
% ========================================================

Let $f(y_i; \mu_i, \phi)$ be the density function for a negative binomial distribution with mean $\mu_i$ and dispersion parameter $\phi$ (hence variance $\mu_i + \phi \mu_i^2$):
\begin{eqnarray}
f(y_i; \mu_i, \phi) = \frac{\Gamma(y_i + 1/\phi)}{y_i! \Gamma(1/\phi)}
\left( \frac{1}{1 + \phi\mu_i} \right)^{1/\phi} \left(\frac{\phi\mu_i}{1 + \phi\mu_i}\right)^{y_i}.
\label{eq:nb}
\end{eqnarray}
As $\phi \rightarrow 0$, $f(y_i; \mu_i, \phi)$ converges to Poisson distribution with mean $\mu_i$. While all the following discussions focus on negative binomial distribution, they can be easily extended to Poisson situation and we omit the details here. Using negative binomial distribution, the log likelihood is
\begin{eqnarray}
l(\mathbf{y}, \boldsymbol{\mu}, \phi) &=&
\sum_{i=1}^n \left[
\log\left( \frac{\Gamma(y_i + 1/\phi)}{y_i!\Gamma(1/\phi)} \right)
+ y_i \log\left(\frac{\phi\mu_i}{1 + \phi\mu_i}\right)
- \frac{1}{\phi} \log(1 + \phi\mu_i) \right],
\end{eqnarray}
where $\mathbf{y} = (x_1, ..., x_n)\trans$, $\boldsymbol{\mu} = (\mu_1, ..., \mu_n)\trans$, and $n$ indicates sample size. We further assume
$\mu_i = \sum_{j=1}^J x_{ij} b_j$, where $b_j \geq 0$, and maximize the penalized log likelihood
\begin{eqnarray}
l(\mathbf{y}, \boldsymbol{\mu}, \phi)  - \sum_{j=1}^p q(b_j),
\label{eqn:pLike}
\end{eqnarray}
where $q(b_j) = \lambda \log(b_j  + \tau)$, and $\lambda$ and $\tau$ are two tuning parameters. In contrast to conventional penalized GLM, we employ a non-canonical link function, does not use an intercept, and impose a set of constraints that $b_j \geq 0$ for $j=1, 2, ..., J$. We maximize the likelihood by iteratively updating regression coefficients $b_j$ and dispersion parameter $\phi$. Following Friedman et al. \cite{Friedman10}, we approximate the likelihood part in equation~(\ref{eqn:pLike}) by a quadratic approximation:
\begin{eqnarray*}
l_Q(\mathbf{y}, \boldsymbol{\mu}, \phi)  = - \sum_{i=1}^n w_i\left( y_i - \sum_{j=1}^p x_{ij}b_j\right)^2,
\end{eqnarray*}
where $w_i = 1/\left(\hat{\mu}_i + \phi \hat{\mu}_i^2\right)$, and $\hat{\mu}_i$ is the estimate of $\mu_i$ in the previous iteration. Then to solve $\bb$, we just need to solve the following penalized least squares problem.
\begin{eqnarray*}
\sum_{i=1}^n w_i\left( y_i - \sum_{j=1}^p x_{ij}b_j\right)^2 + \sum_{j=1}^p q(b_j),
\end{eqnarray*}
subject to the constraints of $b_j \geq 0$ for $1 \leq j \leq J$. We employ a modified Iterative Adaptive Lasso (IAL) algorithm \cite{Sun10} to solve this problem. Given estimates of {$b_j$ ($1 \leq j \leq p$), $\phi$ can be re-estimated by maximizing the conditional likelihood of $\phi$. \\

Specifically, the implementation includes the following four levels of loops:

\begin{itemize}
  \item {\textsc{outer loop}}: Iterate across all combinations of tuning parameter $\lambda_n$ and $\tau_n$.
    \item {\textsc{middle loop 1}}: This corresponds to the loop of iteratively update $b_j$ $(1 \leq j \leq p)$ and $\phi$. For each given $\phi$, we carry out the next two nested loops to estimate $b_j$ and then re-estimate $\phi$.
  \item {\textsc{middle loop 2}}: This corresponds to the loop of IRLS (Iterative Re-weighted Least Squares). Update the quadratic approximation $l_Q$ using current estimate of $b_j$ $(1 \leq j \leq p)$ and $\phi$.
  \item {\textsc{inner loop}}: Run the modified IAL to re-estimate $b_j$ $(1 \leq j \leq p)$ on the penalized weighted least squares problem. \\
\end{itemize}

The modified IAL algorithm is as follows. It is different from the IAL \cite{Sun10} in that the regression coefficients need to be non-negative and we remove the step of estimating residual variance to improve the computational efficiency and robustness. 

\begin{enumerate}
\item \textsc{Initialization}: initialize $b_j$ with zero's or estimate from previous IRLS iteration, and initialize $\kappa_j = (b_j + \tau)/\lambda$, where $1 \leq j \leq p$. 
\item Iterative Updates:
  \begin{enumerate}
  \item For $j = 1, ..., p$, update $b_j$,
    \begin{displaymath}
      \left\{\begin{array}{ll}
        b_j = \bar{b}_j - 1/\kappa_j &
        \textrm{if $\bar{b}_j > 1/\kappa_j$}\\
        b_j = 0 &
        \textrm{otherwise}
      \end{array} \right. ,
    \end{displaymath}
    where
    \begin{eqnarray*}
         \bar{b}_j = \left(\sum_{i=1}^n x_{ij}^2 \right)^{-1}
     \sum_{i=1}^n x_{ij}\left(y_i - \sum_{k \neq j} x_{ik}b_k\right).
     \end{eqnarray*}

    \item Update $\kappa_j$: $\kappa_j = (b_j + \tau)/\lambda$.
    \end{enumerate}
\end{enumerate}
This IAL algorithm is converged if the coefficient estimates $\hat{b}_1, ..., \hat{b}_p$ have little change.\\

Tuning parameter selection is a crucial step for any penalization method. We select the tuning parameters $\lambda$ and $\tau$ by a two-dimensional grid search. By default, we search across 10 values of $\lambda$ and 3 values of $\tau$, which are 30 tuning parameter combinations. Larger $\lambda$ and smaller $\tau$ leads to stronger penalty, and thus we first choose the ratio $\lambda/\tau$'s so that they are uniformly distributed in log scale and the largest $\lambda/\tau$ is large enough to penalize all coefficients to 0. Then $\tau$'s are chosen so that they are uniformly distributed in log scale with largest $\tau$ being 0.1. Finally for each ratio $r=\lambda/\tau$ and for each $\tau$, $\lambda$ can be calculated as $r\tau$. Simulations show that the results are similar if we carry out grid search for 500 or 50 tuning parameter combinations. \\

Through the two-dimensional grid search, we choose the combination of $\lambda$ and $\tau$ that minimizes BIC or extended BIC \cite{chen2012extended} if $n > p$ or $n \leq p$, respectively. If we only study the expression of known RNA isoforms, $p$ is often smaller than $n$ (e.g., see Supplementary Figure 4), and thus BIC is used. In contrast, if we detect de novo RNA isoforms, $p$ is often larger than $n$ (e.g., see Supplementary Figure 5), and thus extended BIC is used. For hypothesis testing of the isoform usage, the rule (BIC or extended BIC) is chosen based on the alternative model and the same rule is applied to the null model. More specifically, BIC is defined as $$\texttt{BIC} = -2 l(\hat{\Theta}) + s \log(n),$$ where $l(\hat{\Theta})$ is log likelihood given parameter estimates $\hat{\Theta}$, $s$ is the number of non-zero coefficients after variable selection, and $n$ is sample size. Following Chen and Chen (2012) \cite{chen2012extended}, the extended BIC is defined as $$\texttt{extBIC} = -2 l(\hat{\Theta}) + s \log(n) + 2 \gamma \log p,$$ where $0 \leq \gamma < 1 - 1/(2\kappa)$ given $p = O(n^\kappa)$. In our simulation and real data studies, since we restrict the number of covariates $p \leq 10n$, we set $\kappa=1$ and choose $\gamma = 1/2$. Tuning parameter selection is an active research area and we do not claim our approach is optimal. However our hypothesis-testing framework rely on parametric bootstrap to resample RNA-seq read counts to calculate p-values. This resampling-based p-value calculation is robust to bias due to suboptimal tuning parameters because any bias that influences the null distribution of the test statistic can be captured through resampling. On the other hand, optimal tuning parameter selection may improve the power of our method. 

% ========================================================
\section{Mouse haploperidol treatment experiment}
% ========================================================

\hspace{3ex}\underline{Ethics Statement.} All animal work was conducted in compliance with the ``Guide for the Care and Use of Laboratory Animals'' (Institute of Laboratory Animal Resources, National Research Council, 1996) and approved by the Institutional Animal Care and Use Committee of the University of North Carolina.\\

\underline{Animals.} The mice used in this study were N=2 inbred C57BL/6J females (one placebo treated, one drug treated) and N=2 (129S1Sv/ImJ x PWK/PhJ)F1 females (one placebo treated, one drug treated). All animals were bred at UNC from mice that were less than 6 generations removed from founders acquired from the Jackson Laboratory (Bar Harbor, ME). Animals were maintained on a 14 hour light, 10 hour dark schedule with lights on at 0600. The housing room was maintained at 20-24C with 40-50\% relative humidity. Mice were housed in standard 20cm $\times$ 30cm ventilated polysulfone cages with laboratory grade Bed-O-Cob bedding. Water and Purina Prolab RMH3000 were available ad libitum. A small section of PVC pipe was present in each cage for enrichment. \\

\underline{Drug treatment.} Slow release haloperidol pellets (3.0 mg/kg/day; Innovative Research of America; Sarasota, FL)\cite{fleischmann2002effect} were implanted subcutaneously with a trocar at 8 weeks of age. Blood plasma was collected via tail nick for a drug concentration assay after 30 days of exposure to haloperidol. Steady-state concentration of haloperidol within the clinically relevant range (10-50 nanomoles/L, nM, or 3.75-19 ng/ml)\cite{hsin2000medication} was achieved for both drug treated animals (C57BL/6J: 19nM, (129S1Sv/ImJ x PWK/PhJ)F1: 24 nM).\\

\underline{Tissue collection.} Mice were sacrificed at 12 weeks of age (following 30 days of drug or placebo treatment) by cervical dislocation without anesthesia to avoid its confounding effects on gene expression. Mice were removed from their home cages at 9:00 AM and sacrificed between 10:00 AM and 12:00 PM. Whole brain was rapidly collected, snap frozen in liquid nitrogen, and pulverized to a fine powder using a BioPulverizer unit (BioSpec Products, Bartlesville, OK).\\

\underline{RNA extraction.} Total RNA was extracted from $\sim$25 mg of tissue powder using automated instrumentation (Maxwell 16 Tissue LEV Total RNA Purification Kit, Promega, Madison, WI). RNA concentration was measured by fluorometry (Qubit 2.0 Fluorometer, Life Technologies Corp., Carlsbad, CA) and RNA quality was verified using a microfluidics platform (Bioanalyzer, Agilent Technologies, Santa Clara, CA). \\

\underline{RNAseq methods.} A multiplex library containing all four samples was prepared using the Illumina (San Diego, CA) TruSeq mRNA Sample Preparation Kit v2 with unique indexed adapters (GCCAAT, ACAGTG, CTTGTA, CAGATC). One microgram of total RNA per sample was used as input and the resulting libraries were quantitated using fluorometry. An Illumina HiSeq 2000 instrument was used to generate 100bp paired-end reads (2x100) in one lane of a flow cell. \\

For the C57BL/6J inbred mice, the mm9 reference was used for alignment. For the F1 animals, we developed a customized RNAseq alignment pipeline tailored to this experiment. Our approach considered these mice as diploid and included two separate alignments that were subsequently merged. This has the advantage of incorporating all known strain-specific genetic variants into the alignment reference sequence to improve alignment quality and to minimize bias caused by differences in genetic distance between the parental genomes to the reference sequence. First, reads from the F1 hybrids were aligned to the appropriate 'pseudogenomes' representing each of the parental genomes using TopHat\cite{Trapnell09} (v1.4, default parameters including segment length 25 bp, 2 mismatches allowed per segment, 2 mismatches total allowed per 100 bp read, and maximum indel of 3 bases). Pseudogenomes are approximations constructed by incorporating all known SNPs and indels into the C57BL/6 genome (mm9)\cite{church2009lineage}. We included all variants reported by a large-scale sequencing effort that included 129S1Sv/ImJ and PWK/PhJ\cite{keane2011mouse} (June 2011 release). Second, we mapped coordinates from the pseudogenome aligned reads to mm9 coordinates. This involved updating the alignment positions and rewriting the CIGAR strings of each aligned read\cite{li2009sequence}. This was necessary as indels alter the pseudogenome coordinates relative to mm9. Third, we annotated each aligned read to indicate the numbers of maternal and paternal alleles (SNPs and indels) observed in a given read and its paired-end mate. Considering the paired-end mates allowed the use of more paired-end reads for ASE. Finally, alignments to maternal and paternal pseudogenomes were merged by computing the proper union of the separate alignments (i.e., the two alignments were combined such that a read aligning to the same position in both alignments was counted once). \\

% ========================================================
\section{Massively Parallel Computing for {\sc IsoDetector}}
% ========================================================

In practice, the penalized estimation in {\sc IsoDetector} is performed on a grid of $\lambda$ and $\tau$ and the best estimate is chosen according to certain model selection criterion such as BIC. For hypothesis testing purpose, this tuning process has to be done on thousands of bootstrap samples for each gene, which incurs formidable computation burden in real applications. Massively parallel computing based on graphical processing units (GPUs) provides a promising solution. However the coordinate descent based algorithm does not particularly suit the massively parallel computing architecture. Here we propose an algorithm based on the minorization-maximization (MM) principle. Like EM algorithm, MM algorithm always increases the objective value and thus is numerically stable. Furthermore, MM algorithm tends to separate variables, making massively parallel computing feasible in high dimensional optimization problems \cite{ZhouLangeSuchard09MM-GPU}.\\

Consider the log likelihood of a negative binomial model with response vector $\y \in \mathbb{N}^n$ and design matrix $\X \in \mathbb{R}^{n \times p}$
\begin{align*}
    \ell (\bb,\phi|\y,\X) &= \sum_{i=1}^n \left[ \log \frac{(\phi^{-1})_{(y_i)}}{y_i!} + y_i \log (\phi \x_i \trans \bb) - (y_i+\phi^{-1}) \log (1+\phi \x_i \trans \bb) \right],
\end{align*}
where $\phi$ is the overdispersion parameter of negative binomial distribution, and $(\phi^{-1})_{(y_i)}$ denotes the rising factorial $\prod_{k=0}^{y_i-1} (\phi^{-1}+k) = \phi^{-1} (\phi^{-1}+1) ... (\phi^{-1} + y_i -1)$. We assume that entries of $\X$ are nonnegative which is true for isoform estimation problem. To simultaneously achieve isoform selection and estimation, {\sc IsoDetector} relies on log penalized estimation due to its attractive properties. In particular we seek to maximize the penalized objective function
\begin{align}
    f(\bb, \phi) = \ell (\bb,\phi \mid \y,\X) - \sum_{j=1}^p \lambda \log (b_j + \tau),  \label{eqn:negbin-log}
\end{align}
where $\lambda$ and $\tau$ are two tuning parameters, subject to the nonnegativity constraint $b_j \ge 0$. \\

The derivation of MM algorithm for maximizing (\ref{eqn:negbin-log}) relies on simple inequalities \cite{ZhouLange10DMMLE}. The strategy is to minorize term by term. By concavity of log function,
\begin{align*}
    \log (\x_i \trans \bb) = \log \left(\sum_j x_{ij} b_j^{(t)}\right) \ge \sum_j w_{ij}^{(t)} \log b_j + c^{(t)},
\end{align*}
where the superscript $t$ indicates iteration number and $c^{(t)}$ is a constant irrelevant to optimization, and
\begin{align*}
    w_{ij}^{(t)} = \frac{x_{ij} b_j^{(t)}}{\sum_j x_{ij} b_j^{(t)}}.
\end{align*}
By the convexity of negative log function, we apply supporting hyperplane inequality to obtain minorizations
\begin{align*}
    - \log (1 + \phi \x_i \trans \bb) &\ge - \frac{\phi \x_i \trans \bb}{1 + \phi \x_i \trans \bb^{(t)}} + c^{(t)}  \\
    - \log(b_j + \tau) &\ge - \frac{b_j}{b_j^{(t)} + \tau} + c^{(t)}.
\end{align*}
Combining above pieces, we obtain an overall minorization function to the objective function (\ref{eqn:negbin-log})
\begin{align*}
    & g(\bb|\bb^{(t)},\phi^{(t)}) \\
    ={}& \sum_{i=1}^n \left[ y_i \sum_j w_{ij}^{(t)} \log b_j - \left(\frac{y_i + (\phi^{(t)})^{-1}}{1+\phi^{(t)} \x_i \trans \bb^{(t)}}\right) \phi^{(t)} \sum_j x_{ij} b_j \right] - \lambda \sum_j \frac{b_j}{b_j^{(t)}+\tau} + c^{(t)}    \\
    ={}& \sum_j \left[ \left( \sum_i y_i w_{ij}^{(t)} \right) \log b_j - \left( \sum_i \frac{x_{ij}(y_i \phi^{(t)} + 1)}{1+\phi^{(t)} \x_i \trans \bb^{(t)}} + \frac{\lambda}{b_j^{(t)}+\tau} \right) b_j \right] + c^{(t)}.
\end{align*}
It is easy to check that 
\begin{align*}
    g(\bb|\bb^{(t)},\phi^{(t)}) &\le f(\bb,\phi) \hspace{.2in} \text{for all } \bb    \\
    g(\bb^{(t)}|\bb^{(t)},\phi^{(t)}) &= f(\bb^{(t)},\phi^{(t)}).
\end{align*}
Therefore maximizing the minorizing function $g(\bb|\bb^{(t)},\phi^{(t)})$ always increases the objective function
\begin{align*}
    f(\bb^{(t+1)},\phi^{(t)}) \ge g(\bb^{(t+1)}|\bb^{(t)},\phi^{(t)}) \ge g(\bb^{(t)}|\bb^{(t)},\phi^{(t)}) = f(\bb^{(t)},\phi^{(t)}).
\end{align*}
Setting derivative of $g(\bb|\bb^{(t)},\phi^{(t)})$ to zero yields an extremely simple update
\begin{align}
    b_j^{(t+1)} = \frac{\sum_i y_i w_{ij}^{(t)}}{\sum_i \frac{x_{ij}(y_i\phi^{(t)} + 1)}{1+\phi^{(t)} \x_i \trans \bb^{(t)}} + \frac{\lambda}{b_j^{(t)}+\tau}}, \hspace{.2in} j=1,\ldots,p,     \label{eqn:MM-update}
\end{align}
involving only trivial algebra. All parameters $b_j$ are separated and thus updated simultaneously, matching perfectly with the massively parallel architecture of GPUs. The nonnegativity constraints $b_j \ge 0$ are also preserved in the update (\ref{eqn:MM-update}). Whenever $b_j^{(0)}$ are positive, all subsequent iterates $b_j^{(t)}$ will always be nonnegative. Effects of the tuning parameters $\lambda$ and $\tau$ are clear: large $\lambda$ and small $\tau$ cause more shrinkage and vice versa. Furthermore, the log penalty penalizes small $b_j$ more heavily than large $b_j$, a desired property the lasso penalty lacks. \\

The update (\ref{eqn:MM-update}) always increases the objective value. However, it does not update the overdispersion parameter $\phi$. In practice, we update $\phi$ after every a few (e.g., five) updates of $\bb$ by (\ref{eqn:MM-update}). Updating of $\phi$ can be done by either Newton's steps or by invoking MM algorithm again. Both are simple because it is a smooth univariate optimization problem. For brevity the details are omitted here.

%In a numerical experiment in \cite{ZhouLangeSuchard09MM-GPU}, a similar MM algorithm for PET imaging problem shows up to 50 fold speed up on an already outdated GPUs compared to CPU code. Therefore we are optimistic.
\clearpage

% ========================================================
\section{Supplementary Figures}
% ========================================================

\begin{figure}[htbp]
  \centering
\includegraphics[width=6in]{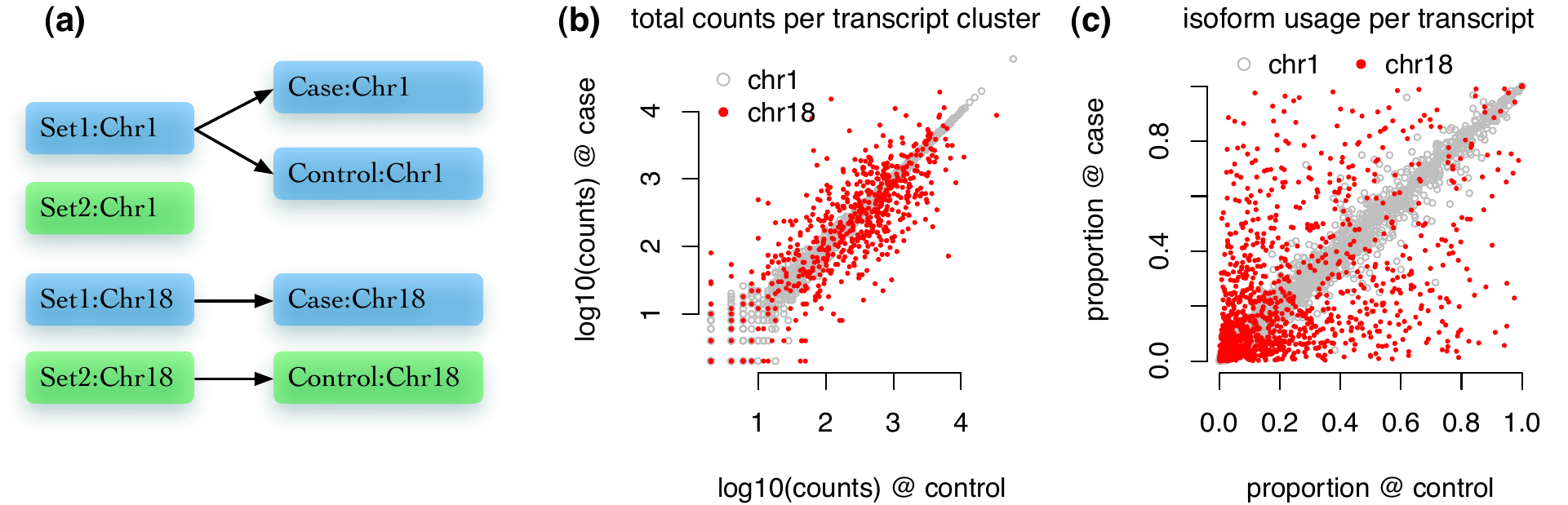}
\caption{ A summary of simulated data. We first simulated $\sim$2 million 76+76bps paired-end reads for data set 1 and data set 2, based on the transcriptome annotation of chromosome 1 and 18 of mouse genome. The expression of any gene/transcript are independent between data set 1 and data set 2. Then as illustrated in \textbf{(a)}, a case and a control sample were generated as follows. For chromosome 1, the sequence fragments of simulation set 1 were randomly split into the case and control samples. For chromosome 18, half of the sequence fragments from set 1 were selected for case and half of the sequence fragments from set 2 were selected for control. Therefore, comparing case and control, all the transcripts in chromosome 1 were equivalently expressed and all the transcripts in chromosome 18 were differentially expressed, either in terms of total expression \textbf{(b)} or isoform usage \textbf{(c)}.  \textbf{(a)} Comparison of the total number of fragments per transcript cluster between the case and the control samples. \textbf{(c)} Comparison of isoform usage of each transcript between the case and the control samples. Here isoform usage is quantified by the ratio of the number of sequence fragments of one transcript over the total number of fragments of the corresponding transcript cluster.}
\end{figure}

\begin{figure}[htbp]
  \centering
\includegraphics[width=5in]{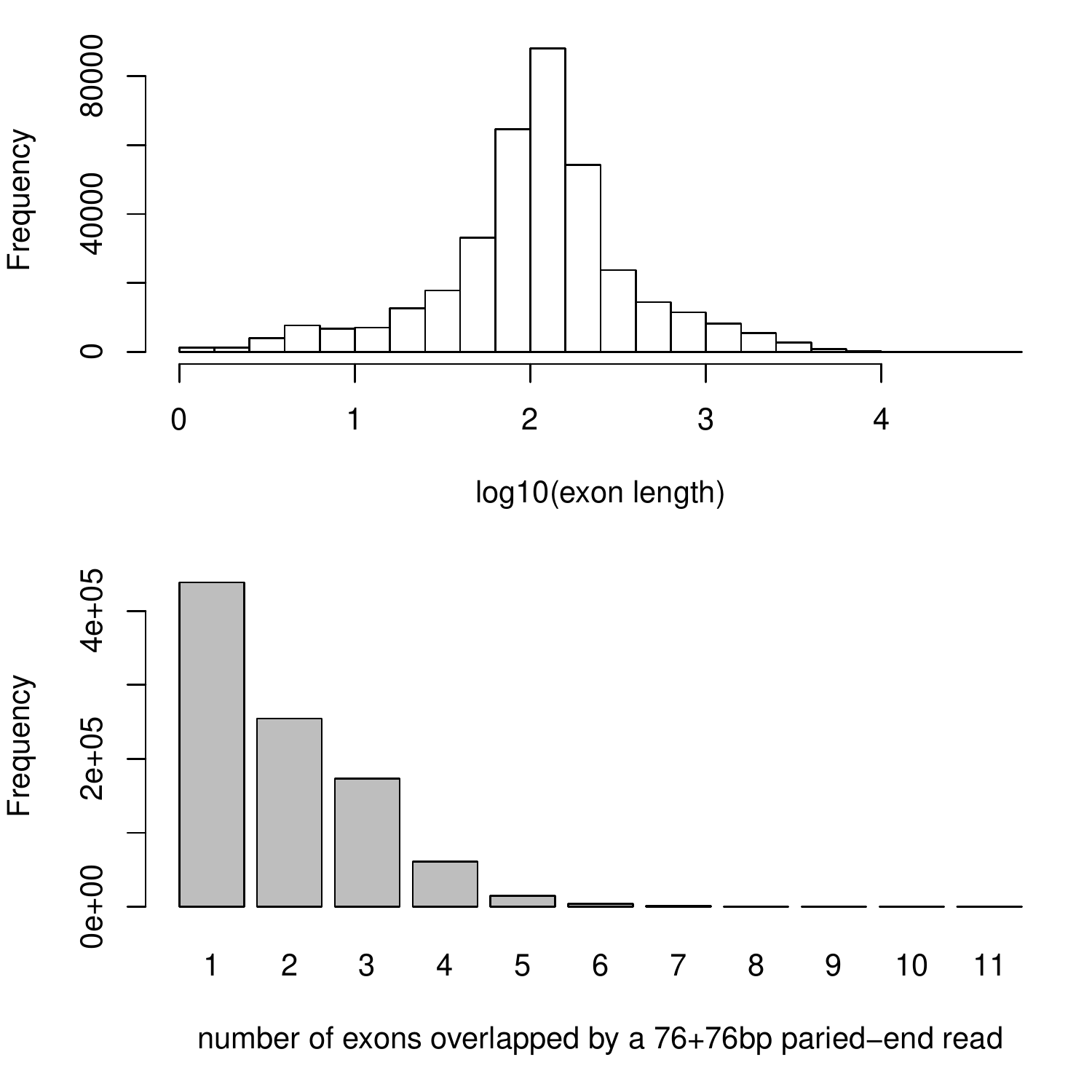}
\caption{ Upper panel shows the distribution of the lengths of non-overlapping exons. Lower panel shows the distribution of the number of exons overlapped by each paired end read. A paired-end read overlaps an exon if at least one base pair of either end of the read overlap with the exon.  About 46\%, 27\%, and 18\% of the reads overlap with only one, two, or three exons. }
\end{figure}

\begin{figure}[htbp]
  \centering
\includegraphics[width=2.5in]{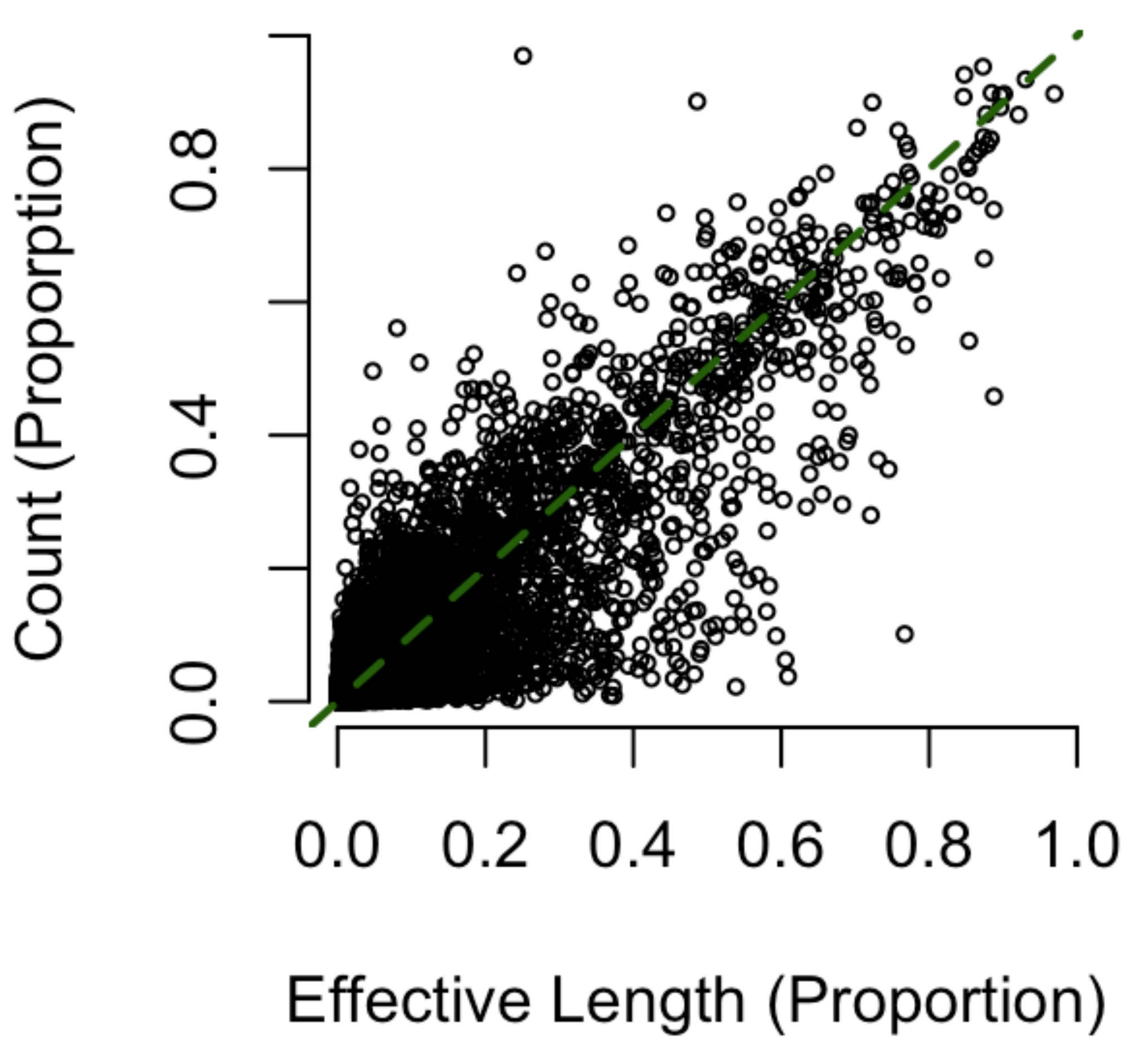}
\caption{ The relation between the effective length of an exon set (divided by the total effective length of the transcript cluster to which the exon belongs) and the proportion of RNA-seq fragments mapped to this exon set in our simulated data. The correlation between them is 0.88. Because different transcript clusters have different expression abundance, we compared read count and effective length as the proportions over the corresponding transcript cluster. Note that the effective length is calculated while assuming all the exons in an exon set are contiguous, which may not be true. Therefore the results here can only be viewed as an approximation.  }
\end{figure}

\begin{figure}[htbp]
  \centering
\includegraphics[width=4.5in]{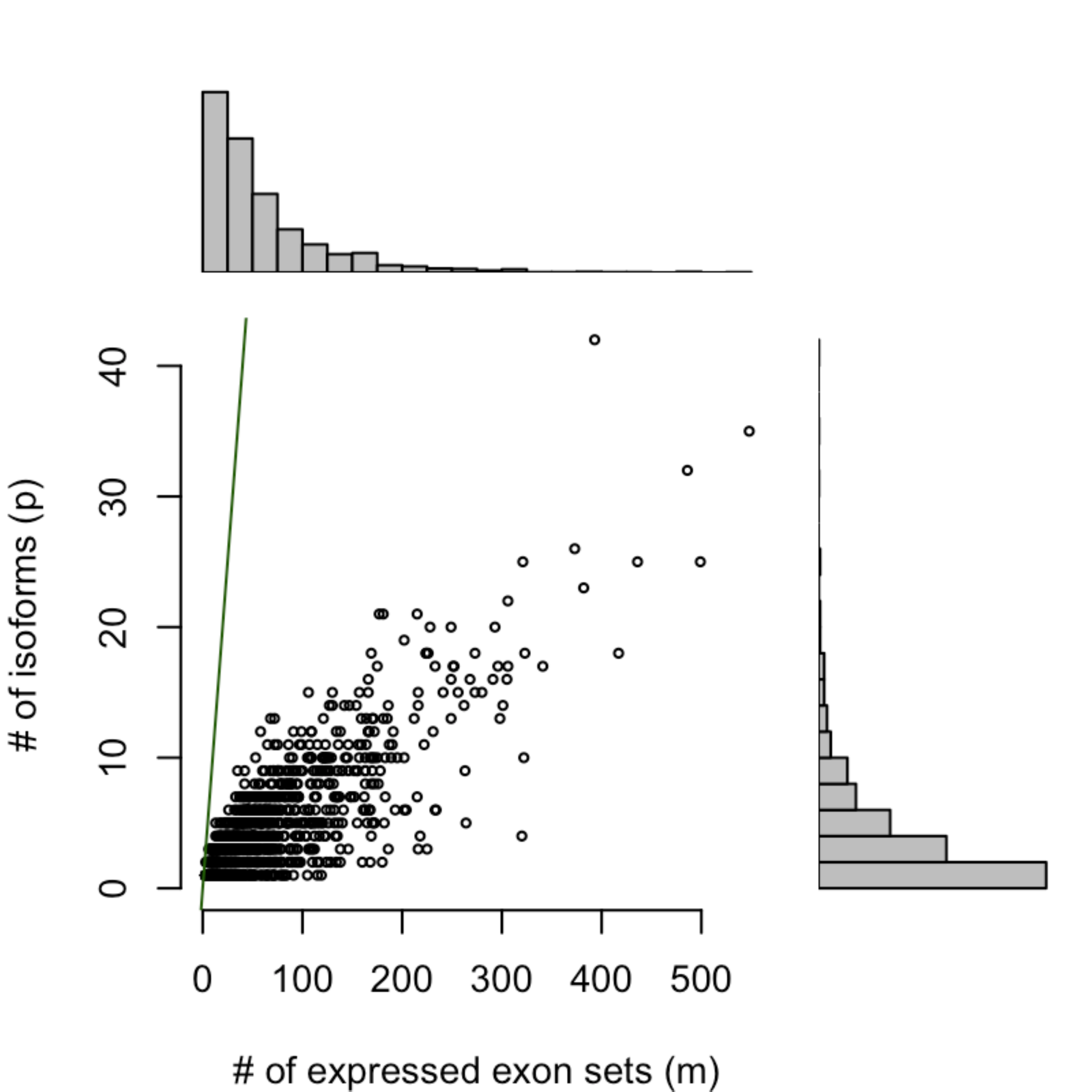}
\caption{ An illustration of the dimension of the isoform selection problem when we use known transcriptome annotation. For each transcript cluster, we consider a variable selection problem where sample size $n$ is the number of expressed exon sets, and the number of covariates $p$ is the number of (candidate) isoforms. The solid line indicates $p=n$. }
\end{figure}

\begin{figure}[htbp]
  \centering
\includegraphics[width=4.5in]{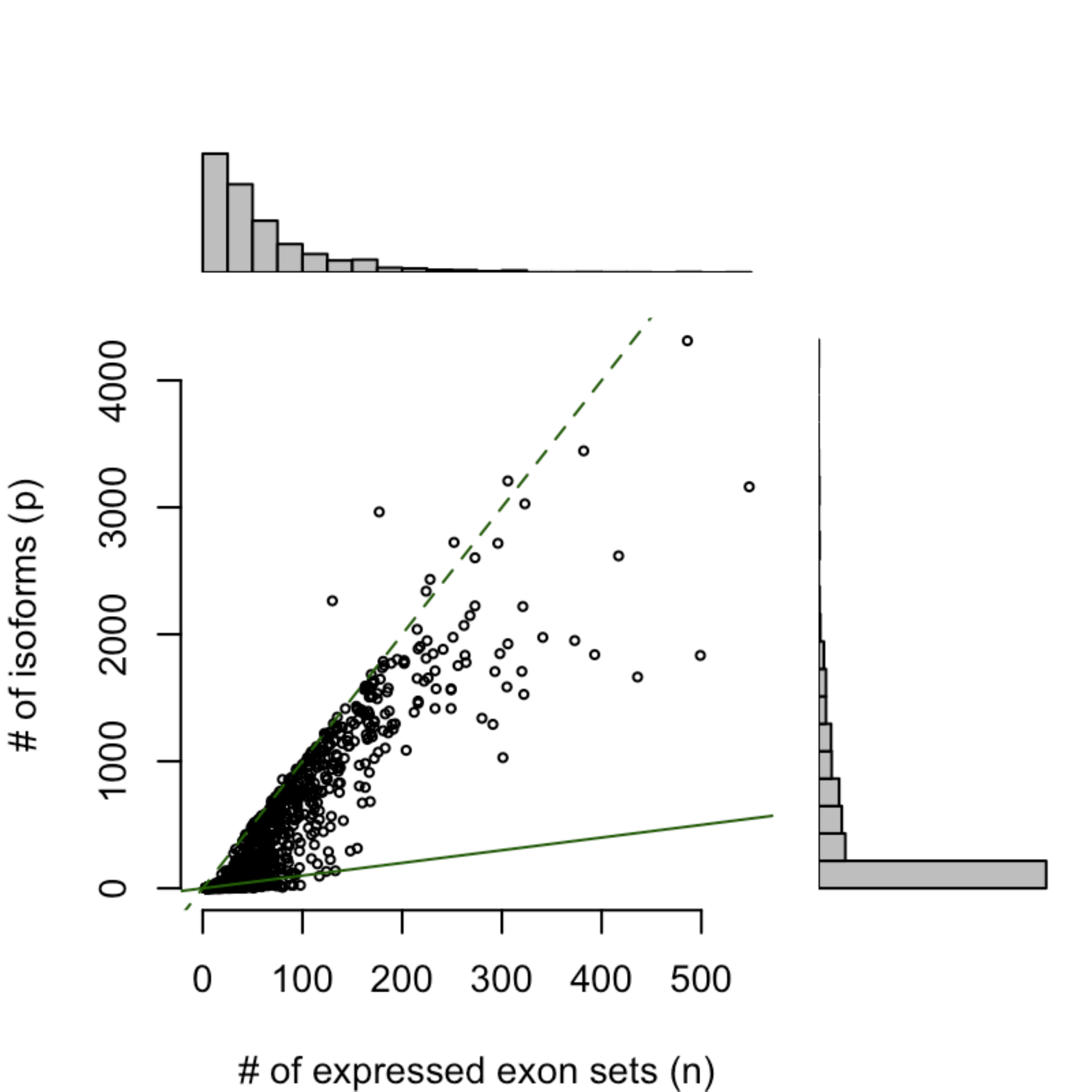}
\caption{ An illustration of the dimension of the isoform selection problem when there is no isoform annotation. For each gene (or transcript cluster), we consider a variable selection problem where sample size $n$ is the number of expressed exon sets, and the number of covariates $p$ is the number of (candidate) isoforms. The solid line indicates $p=n$, and the broken line indicates $p=10n$. In our implementation, we choose the number of candidate isoforms so that $p \leq 10 n$ approximately. Users can loose this restriction with price of increasing computational cost. Our experience is that IsoDetector runs well for $p \leq 100n$. }
\end{figure}

\begin{figure}[htbp]
  \centering
\includegraphics[width=4.5in]{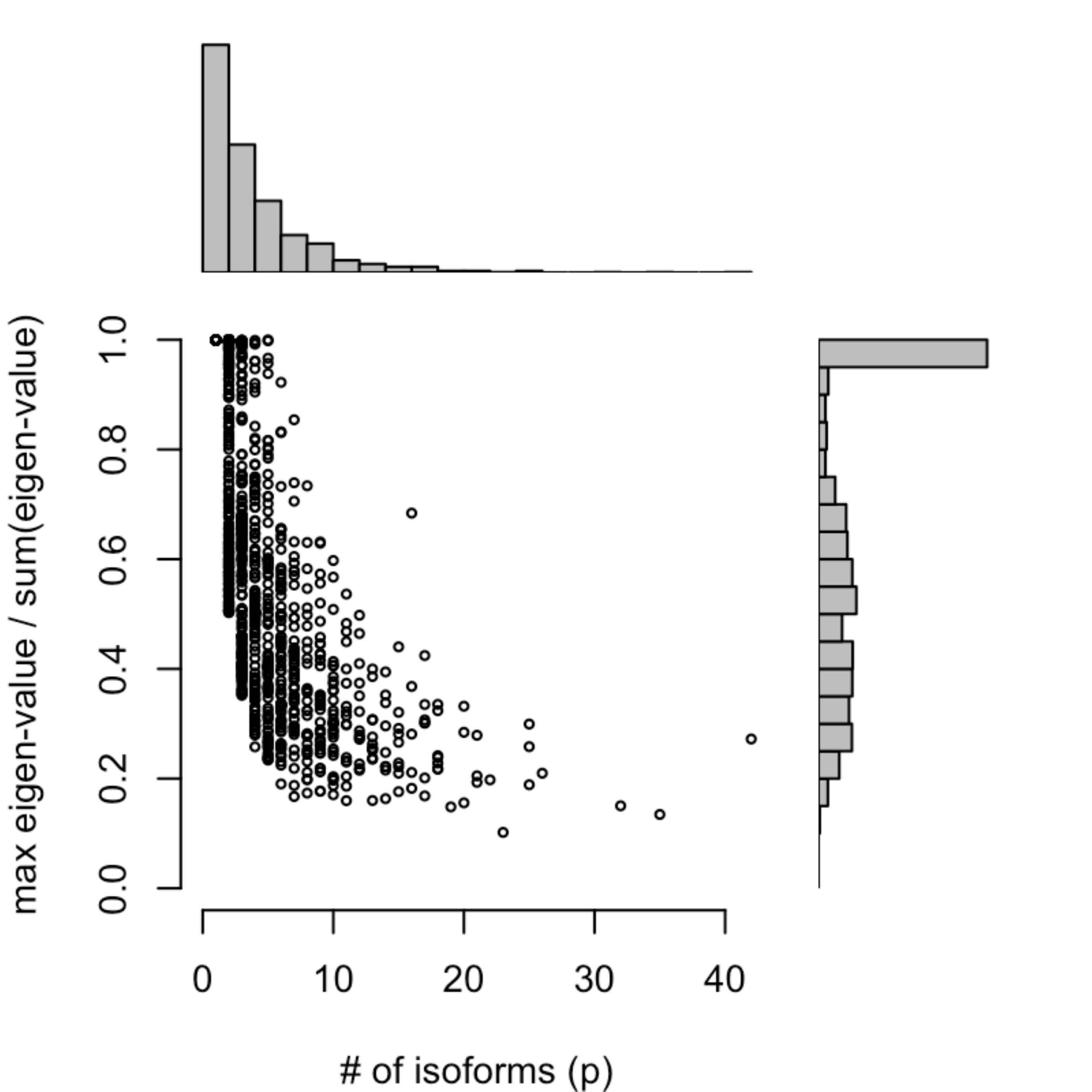}
\caption{ An illustration of the correlation among the isoforms of each transcript cluster when we use known isoform annotation. Each point indicates a transcript cluster where x-axis is the number of isoforms and y-axis is the proportion of variance explained by the first principal component. }
\end{figure}

\begin{figure}[htbp]
  \centering
\includegraphics[width=4.5in]{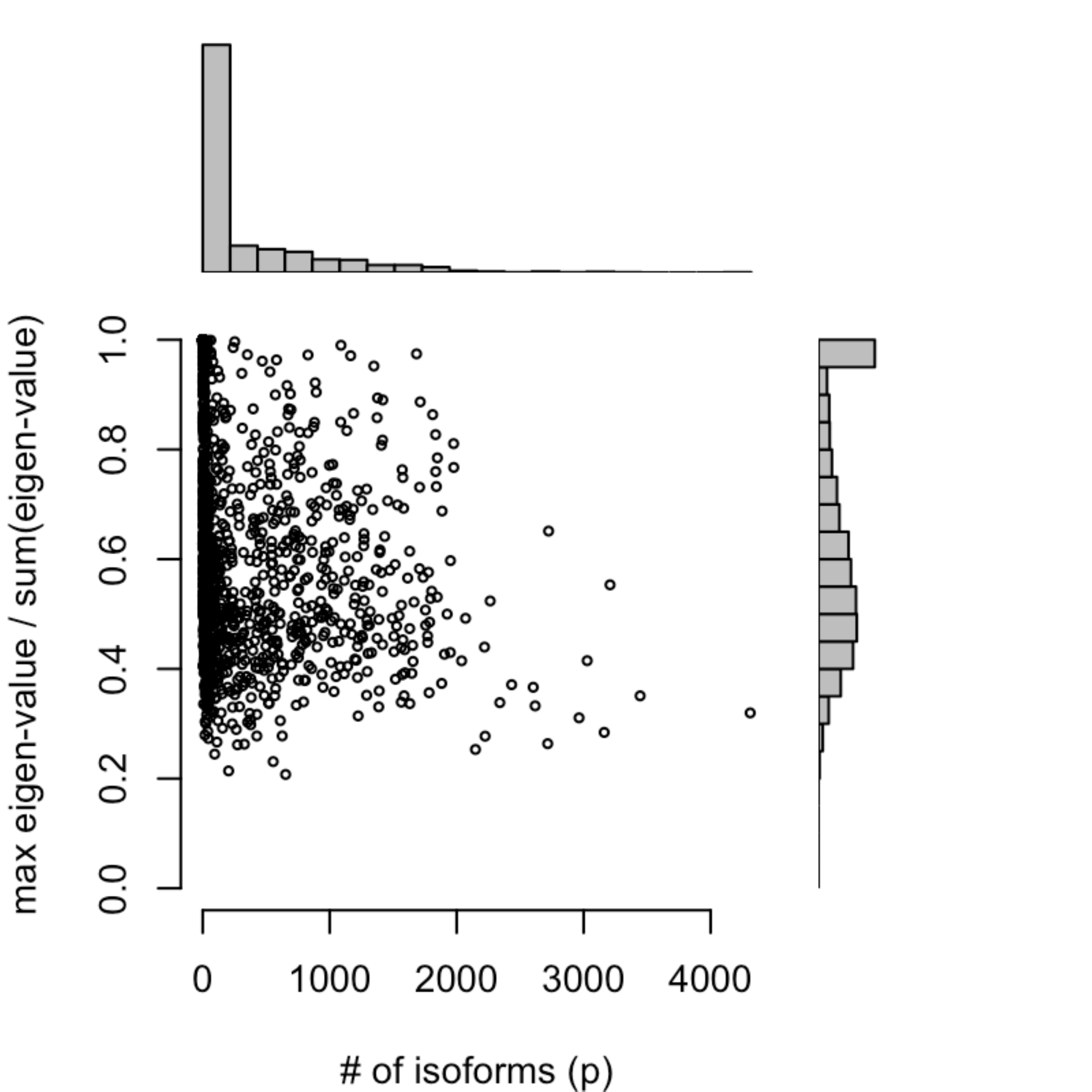}
\caption{ An illustration of the correlation among the isoforms of each transcript cluster when there is no isoform annotation. Each point indicates a transcript cluster where x-axis is the number of isoforms and y-axis is the proportion of variance explained by the first principal component. }
\end{figure}

\begin{figure}[htbp]
  \centering
\includegraphics[width=4.5in]{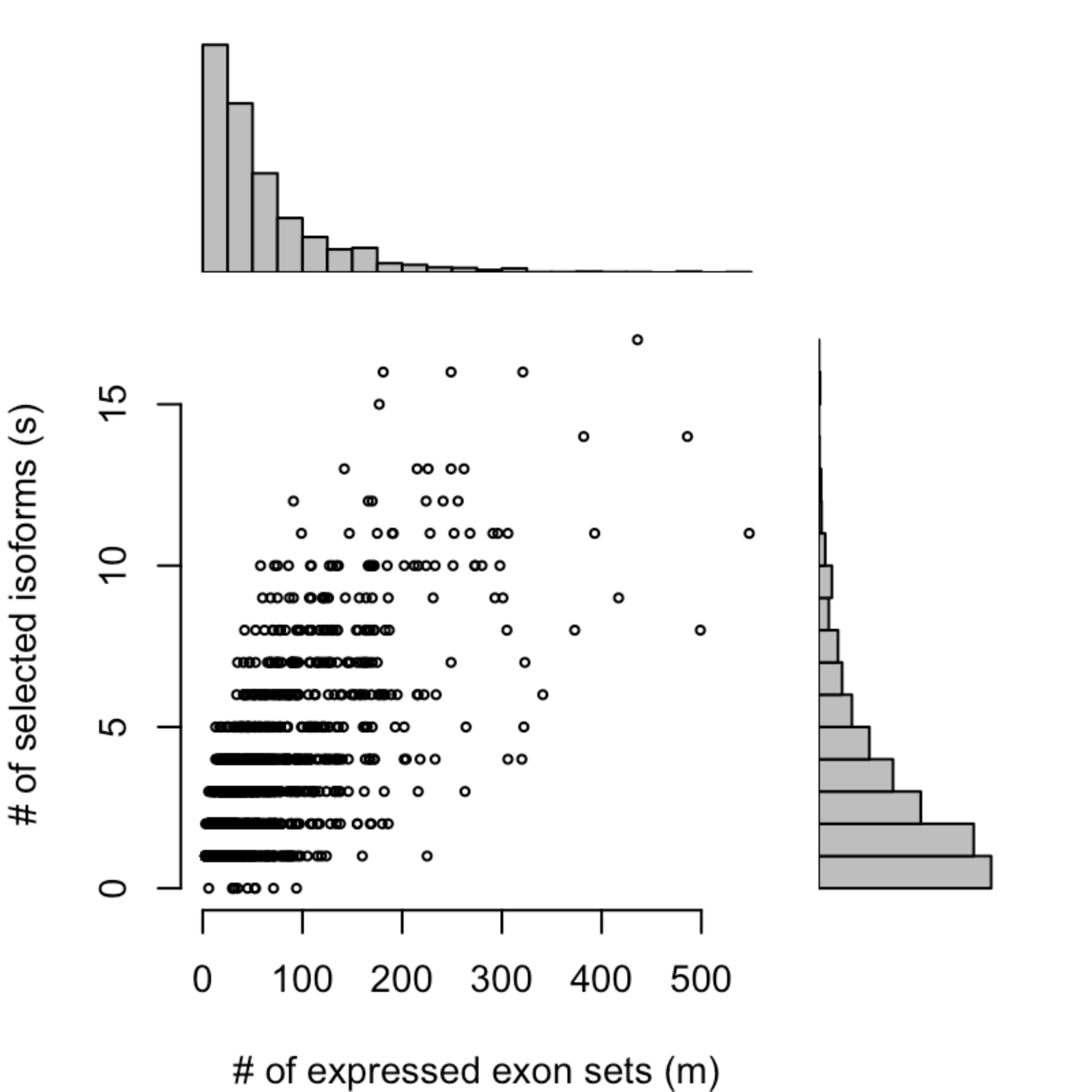}
\caption{ An illustration of the number of isoforms selected by IsoDetector when we choose the candidate isoforms based on the known isoform annotation. }
\end{figure}

\begin{figure}[htbp]
  \centering
\includegraphics[width=4.5in]{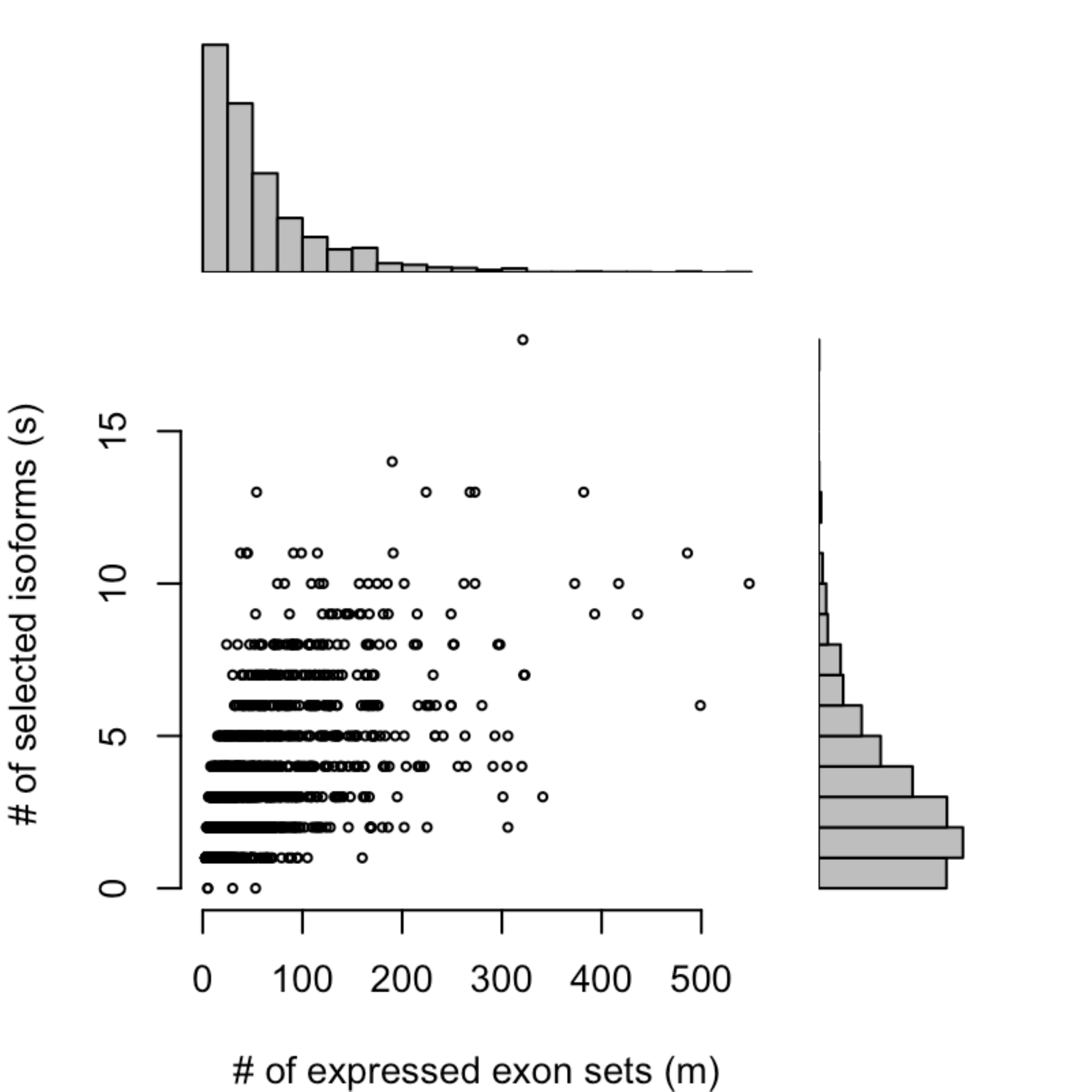}
\caption{ An illustration of the number of isoforms selected by IsoDetector when we choose the candidate isoforms without using any isoform annotation. }
\end{figure}

\begin{figure}[htbp]
  \centering
\includegraphics[width=4.55in]{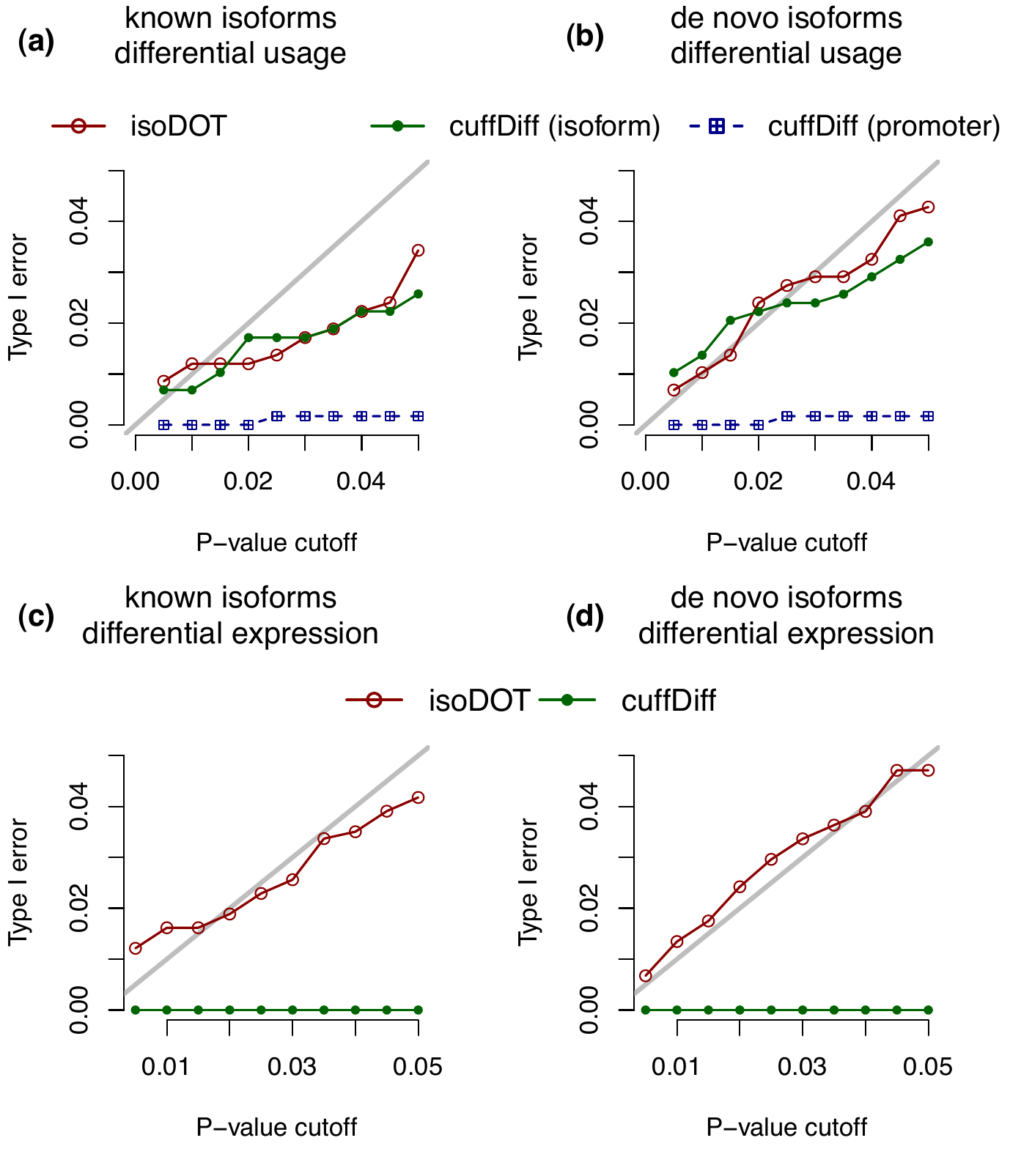}
\caption{ Compare the type I error of IsoDOT and Cuffdiff2 for detesting genes with differential isoform usage (a-b) or differential expression (c-d), while transcriptome annotation is known (a,c) or not (b,d). For the case of differential isoform usage, cufflinks provides results for ``isoform'' and ``promoter'', where the former is for isoform sharing a TSS, and the latter is for differential usage of TSSs. For the ``isoform'' case, we have collapsed the p-values of multiple tests of a gene by taking minimum, thus it leads to an over-estimate of type I error. }
\end{figure}

\begin{figure}[htbp]
  \centering
\includegraphics[width=5in]{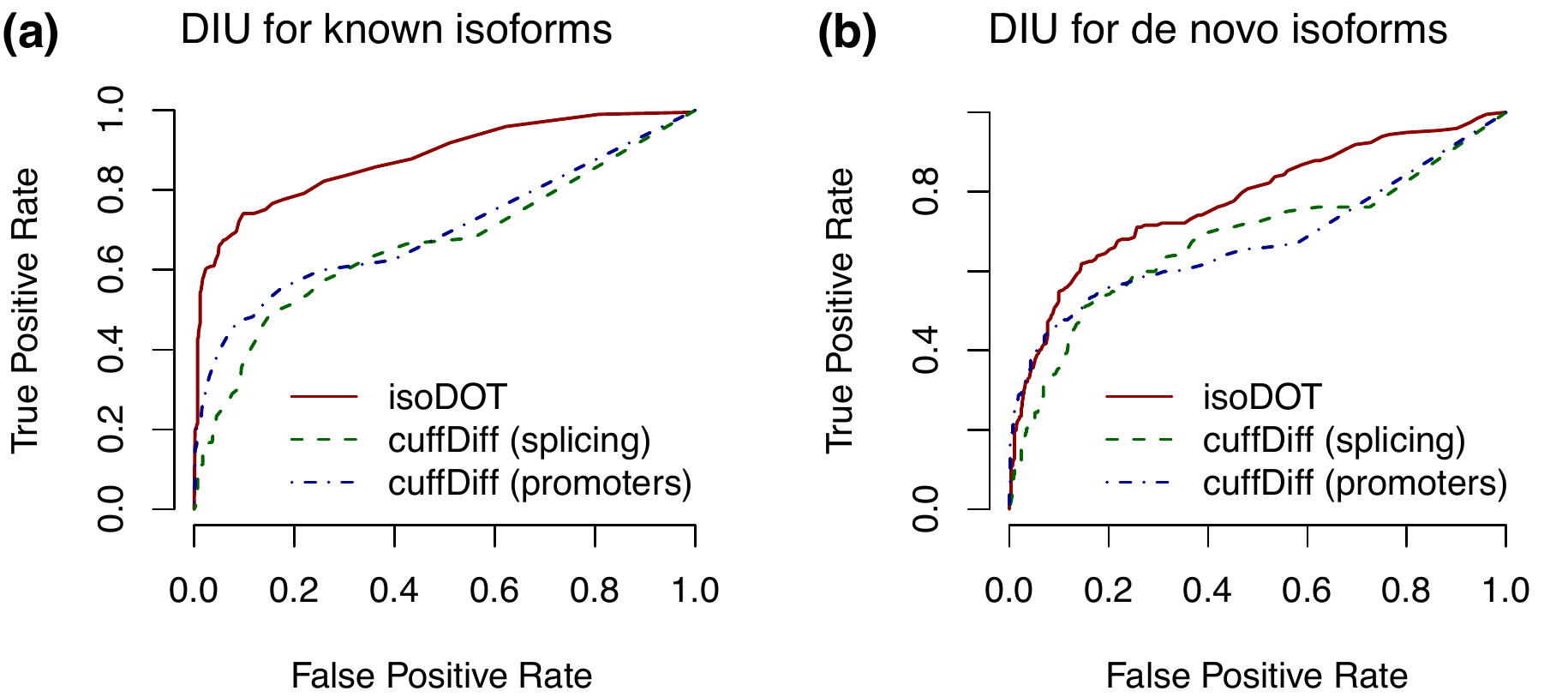}
\caption{ Compare the results of IsoDOT and Cuffdiff2 by ROC curves. The ROC curves compare two methods across a wide range of p-value cutoffs. If one method has a calibration issue (e.g., p-value is larger than it should be) but still ranks the genes correctly, it would perform well judged by ROC curve. The results shown here demonstrate that the IsoDOT still performs better than Cuffdiff2 even if we allow the results of Cuffdiff2 to be calibrated.}
\end{figure}

\begin{figure}[htbp]
  \centering
\includegraphics[width=6.4in]{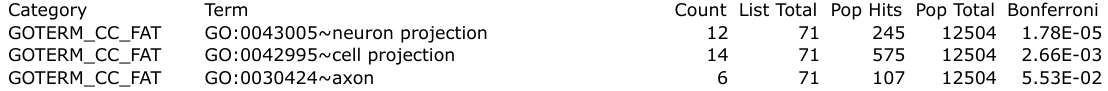}
\caption{ DAVID functional category enrichments for DIU genes (with transcriptome annotation). }
\end{figure}

\begin{figure}[htbp]
  \centering
\includegraphics[width=6.4in]{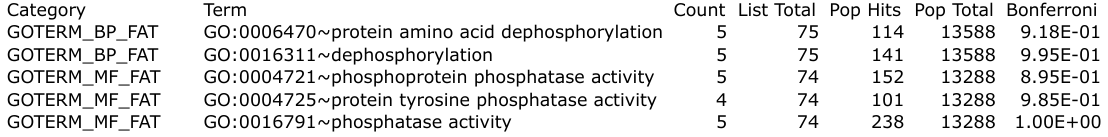}
\caption{ DAVID functional category enrichments for DIE genes (with transcriptome annotation). }
\end{figure}

\begin{figure}[htbp]
  \centering
\includegraphics[width=6.4in]{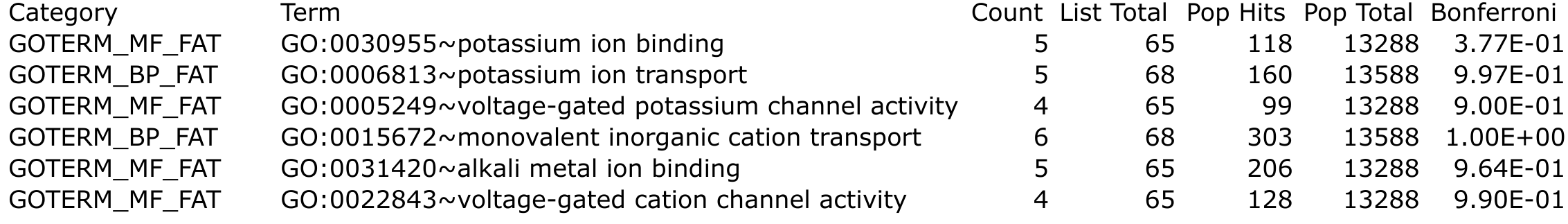}
\caption{ DAVID functional category enrichments for DIU genes (without  transcriptome annotation). }
\end{figure}

\begin{figure}[htbp]
  \centering
\includegraphics[width=6.4in]{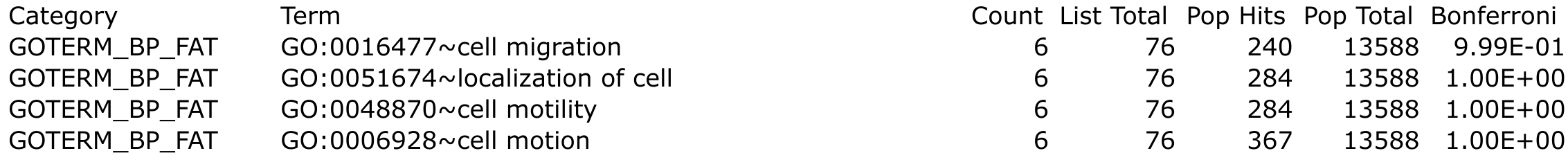}
\caption{ DAVID functional category enrichments for DIE genes (without  transcriptome annotation). }
\end{figure}

\begin{figure}[htbp]
  \centering
\includegraphics[width=6.4in]{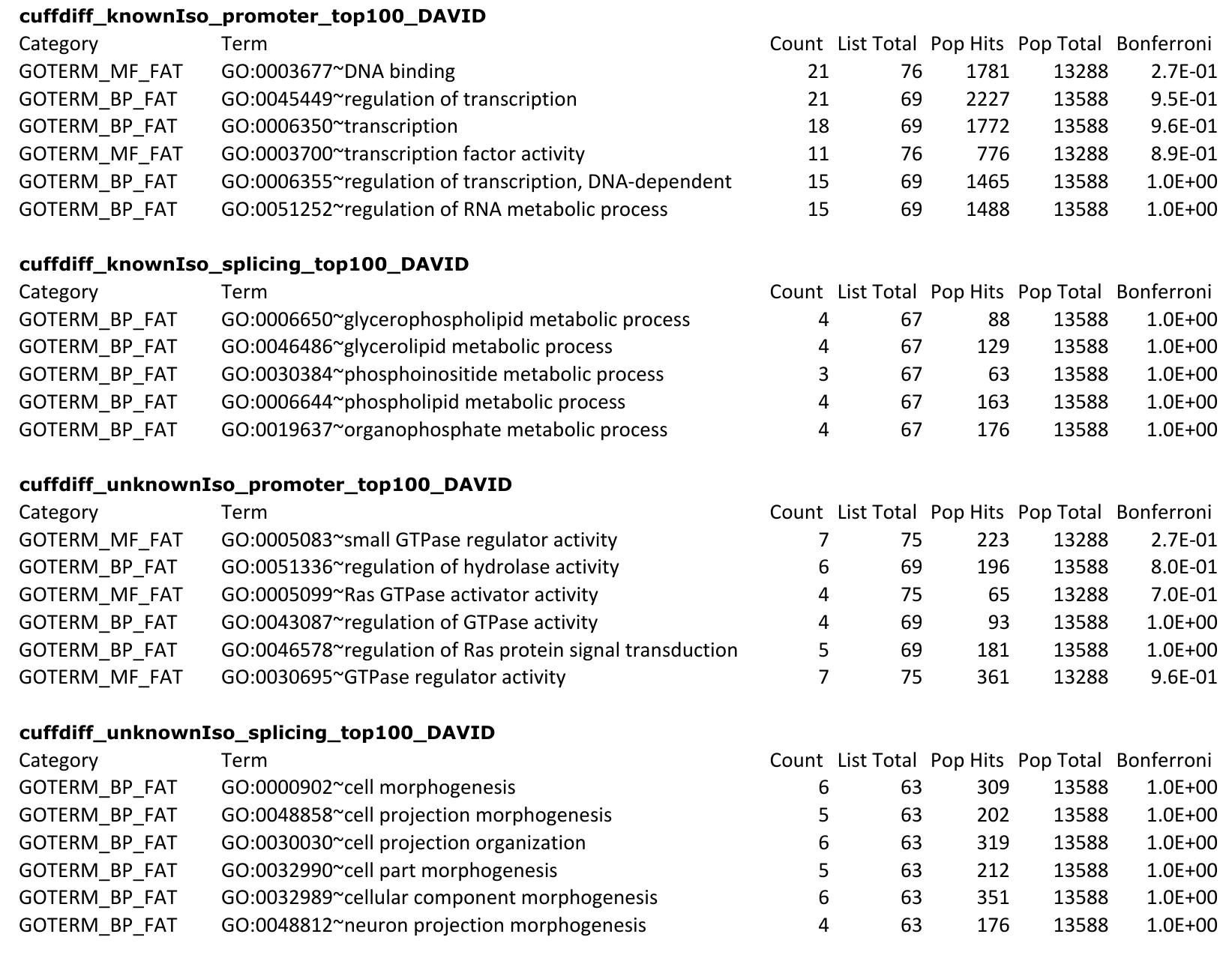}
\caption{ DAVID functional category enrichments for the results from Cuffldiff. }
\end{figure}

\begin{figure}[htbp]
  \centering
\includegraphics[width=4.5in]{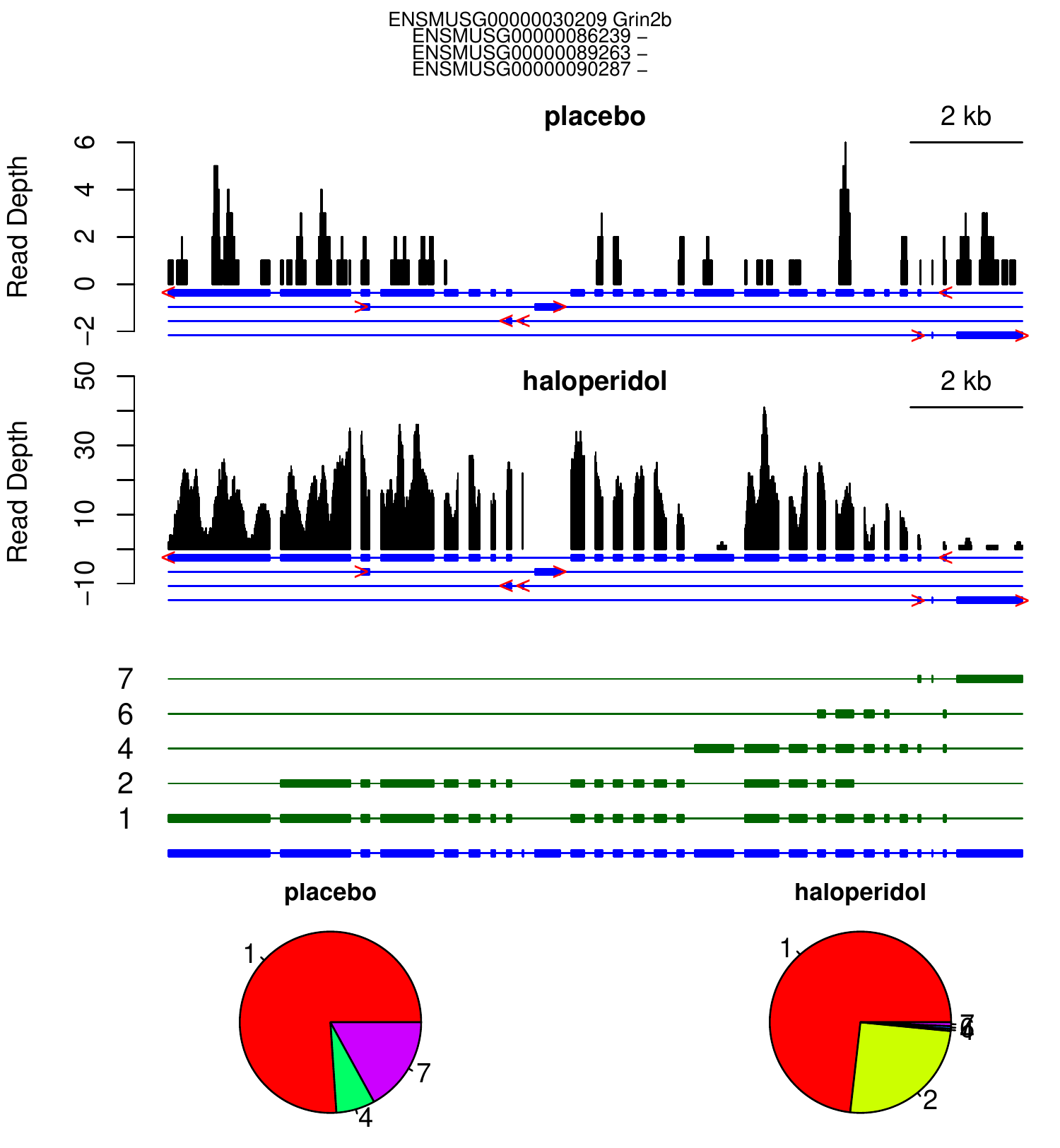}
\caption{ Differential isoform usage of gene Grin2b between two C57BL/6 mice with haloperidal or placebo treatment. Note Grin2b belongs to a transcript cluster with four genes. However, the other three genes are short and contribute little if any signal of differential isoform usage.}
\end{figure}

\begin{figure}[htbp]
  \centering
\includegraphics[width=2.5in]{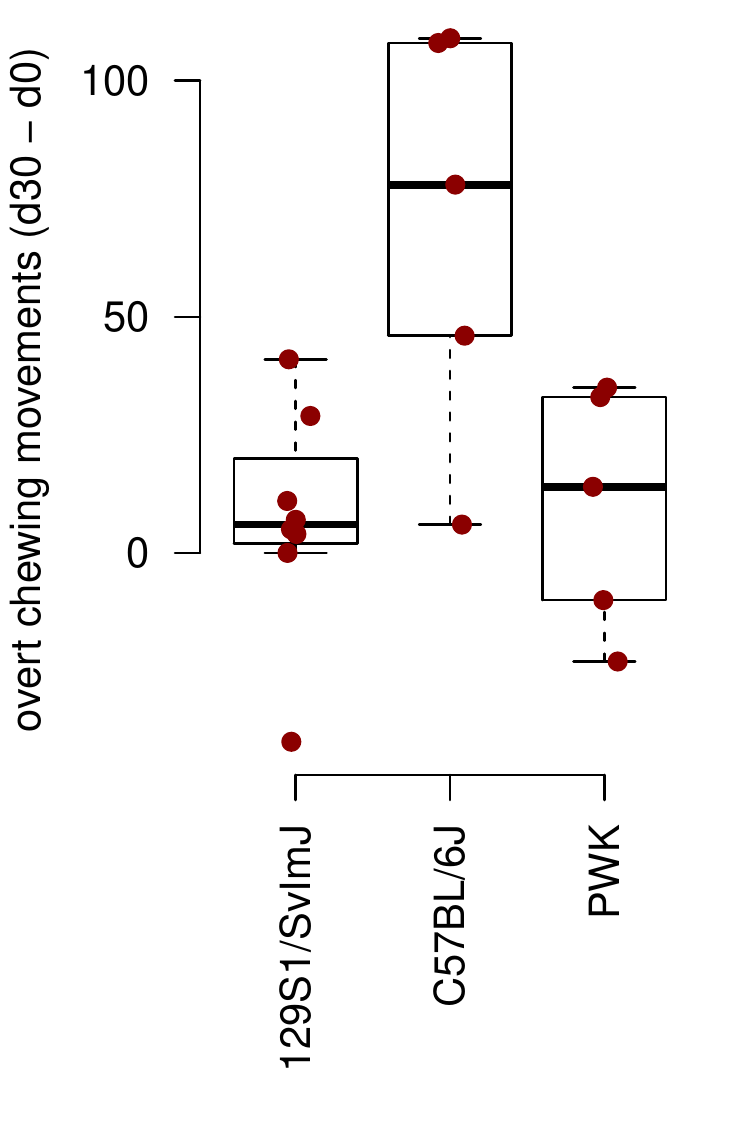}
\caption{Change of overt chewing movements for three inbred strains between day 0 and day 30 after haloperidol treatment. See Crowley et al. \cite{crowley2012antipsychotic} for more details of the experiment and the results of other phenotypic outcomes. }
\end{figure}

\clearpage

\begin{figure}[htbp]
  \centering
\includegraphics[width=3.5in]{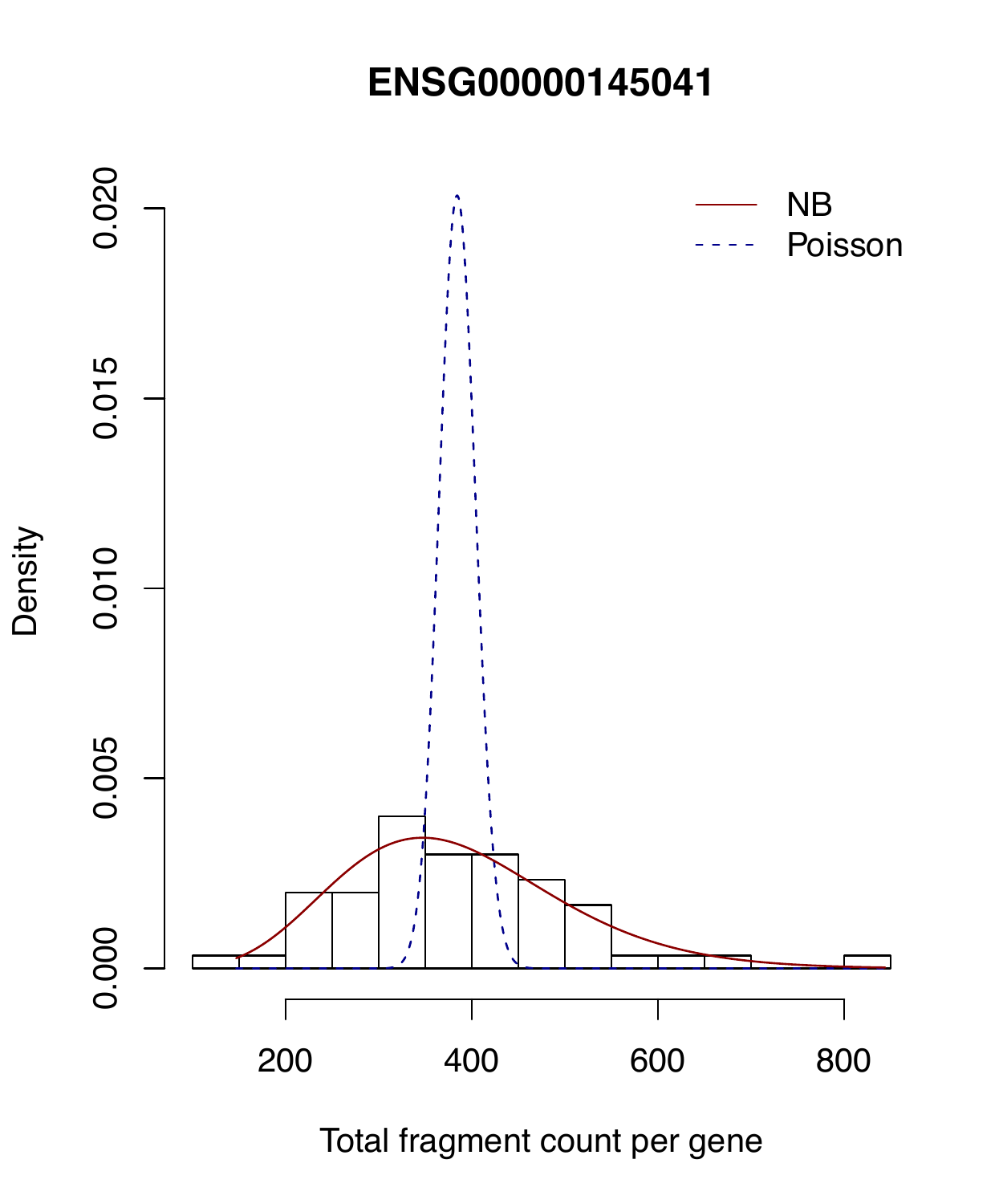}
\caption{An example to show that negative binomial distribution can provide adequate fit to RNA-seq fragment count data whereas Poisson distribution assumption leads to severe underestimate of variance. The RNA-seq data used in this example are the RNA-seq fragment count for gene VPRBP (Vpr (HIV-1) binding protein, ensembl ID: ENSG00000145041) from 50 HapMap CEU samples \cite{Montgomery10}. The MLE of the two distributions were obtained using R function glm and glm.nb, respectively, after correction for read-depth.}
\end{figure}

\begin{figure}[htbp]
  \centering
\includegraphics[width=6.5in]{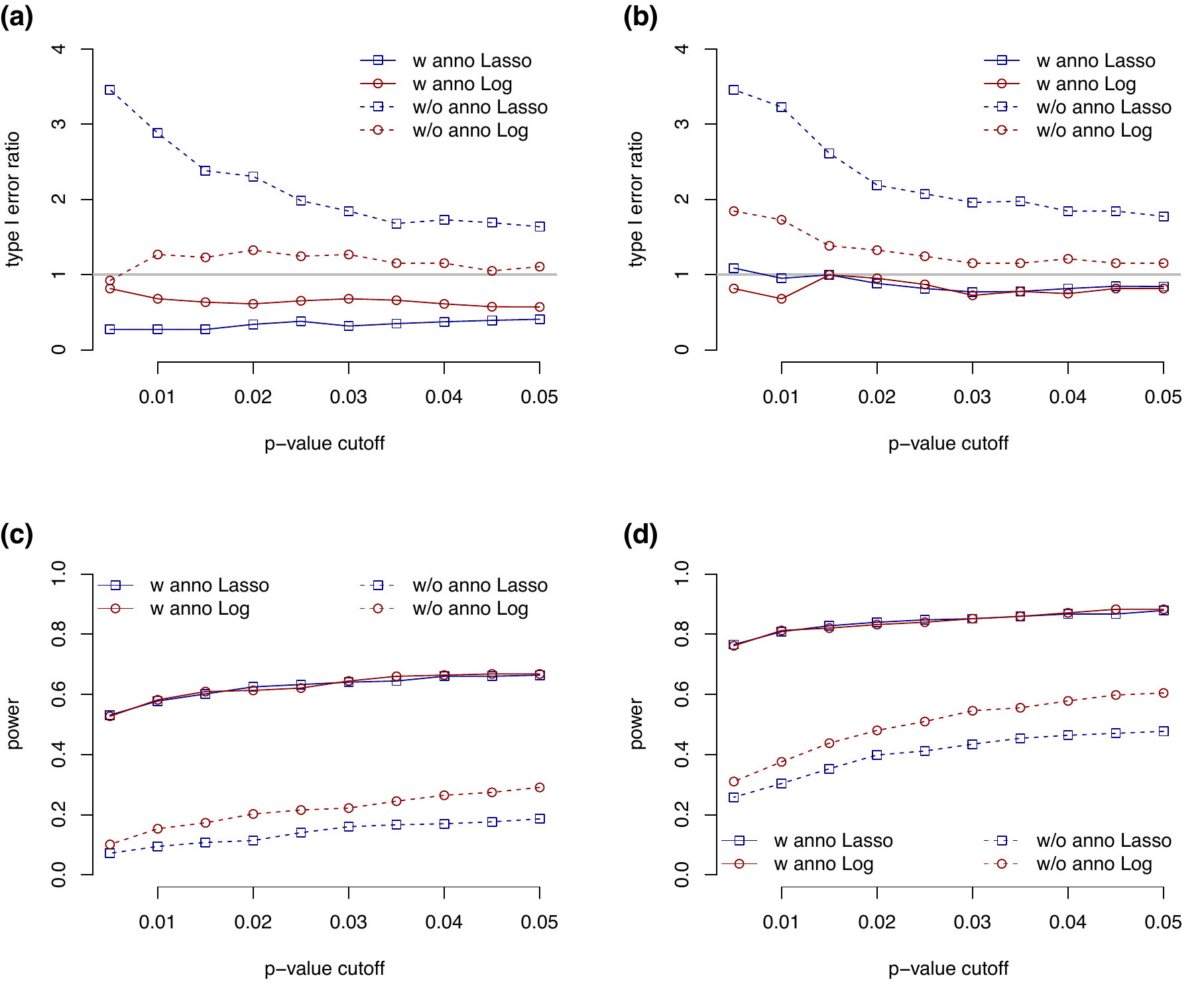}
\caption{Compare type I error and power when we use the Lasso penalty or the Log penalty in IsoDOT. (a) Type I for DIU test. (b) Type I error for DIE test. (c) Power for DIU test. (d) Power for DIE test. In panel (a) and (b), the y-axis is type I error ratio, which is the ratio of observed type I error rate divided by the corresponding p-value cutoff (x-axis), which is the expected type I error rate. }
\end{figure}

\clearpage

\begin{table}[htbp]
\begin{center}
  \caption{Read depth of the RNA-seq data in mouse haloperidol treatment study. Each sequence fragment was sequenced on both ends by 93-100bp. The first four rows show the information of four mince and the last four rows are for allele-specific RNA-seq reads from the two F1 mice. }
  \vspace{1ex}
\begin{tabular}{l l l l l }
  \hline                        
 Sample & Genetic & Treatment & Total number & Number of fragments \\
 ID & background & & of mapped reads & passed QC and mapped\\
     & & & & to exonic regions\\
  \hline                        
BB1050 &  C57BL/6J & placebo & 21,482,924 & 8,337,872 \\
BB1068 &  C57BL/6J & haloperidol & 27,178,749 & 10,486,170 \\
CG0069 & 129$\times$PWK & placebo & 24,014,041 & 10,476,460 \\
CG0077 & 129$\times$PWK & haloperidol & 20,365,336 & 8,871,864\\ 
  \hline                        
CG0069 & 129 @ 129$\times$PWK & placebo & 4,667,545 & 1,953,335 \\
CG0069 & PWK @ 129$\times$PWK & placebo & 4,605,879 & 1,931,791\\ 
CG0077 &  129 @ 129$\times$PWK & haloperidol & 3,993,348 & 1,668,243 \\
CG0077 &  PWK @ 129$\times$PWK & haloperidol & 3,957,371 & 1,654,705 \\
  \hline                        
\end{tabular}
\end{center}
\end{table}

\begin{center}
\begin{longtable}{l l p{9.3cm}}
\caption{Top 100 genes identified from differential isoform usage (DU only) analysis comparing two C57BL/6J mice with haloperidol or placebo treatments. }\\
\hline
\textbf{Ensembl ID} & \textbf{symbol} & \textbf{name} \\
\hline
\endfirsthead
\multicolumn{3}{c}%
{\tablename\ \thetable\ -- \textit{Continued from previous page}} \\
\hline
\textbf{Ensembl ID} & \textbf{symbol} & \textbf{name} \\
\hline
\endhead
\hline \multicolumn{3}{r}{\textit{Continued on next page}} \\
\endfoot
\hline
\endlastfoot
ENSMUSG00000040537 & Adam22 & a disintegrin and metallopeptidase domain 22\\
ENSMUSG00000020431 & Adcy1 & adenylate cyclase 1\\
ENSMUSG00000049470 & Aff4 & AF4/FMR2 family, member 4\\
ENSMUSG00000061603 & Akap6 & A kinase (PRKA) anchor protein 6\\
ENSMUSG00000040407 & Akap9 & A kinase (PRKA) anchor protein (yotiao) 9\\
ENSMUSG00000069601 & Ank3 & ankyrin 3, epithelial\\
ENSMUSG00000071176 & Arhgef10 & Rho guanine nucleotide exchange factor (GEF) 10\\
ENSMUSG00000059495 & Arhgef12 & Rho guanine nucleotide exchange factor (GEF) 12\\
ENSMUSG00000002343 & Armc6 & armadillo repeat containing 6\\
ENSMUSG00000020788 & Atp2a3 & ATPase, Ca++ transporting, ubiquitous\\
ENSMUSG00000003604 & Aven & apoptosis, caspase activation inhibitor\\
ENSMUSG00000048251 & Bcl11b & B-cell leukemia/lymphoma 11B\\
ENSMUSG00000049658 & Bdp1 & B double prime 1, subunit of RNA polymerase III transcription initiation factor IIIB\\
ENSMUSG00000042460 & C1galt1 & core 1 synthase, glycoprotein-N-acetylgalactosamine 3-beta-galactosyltransferase, 1\\
ENSMUSG00000039983 & Ccdc32 & coiled-coil domain containing 32\\
ENSMUSG00000033671 & Cep350 & centrosomal protein 350\\
ENSMUSG00000021097 & Clmn & calmin\\
ENSMUSG00000060924 & Csmd1 & CUB and Sushi multiple domains 1\\
ENSMUSG00000048796 & Cyb561d1 & cytochrome b-561 domain containing 1\\
ENSMUSG00000017999 & Ddx27 & DEAD (Asp-Glu-Ala-Asp) box polypeptide 27\\
ENSMUSG00000037426 & Depdc5 & DEP domain containing 5\\
ENSMUSG00000024456 & Diap1 & diaphanous homolog 1 (Drosophila)\\
ENSMUSG00000045103 & Dmd & dystrophin, muscular dystrophy\\
ENSMUSG00000041268 & Dmxl2 & Dmx-like 2\\
ENSMUSG00000039716 & Dock3 & dedicator of cyto-kinesis 3\\
ENSMUSG00000036270 & Edc4 & enhancer of mRNA decapping 4\\
ENSMUSG00000028760 & Eif4g3 & eukaryotic translation initiation factor 4 gamma, 3\\
ENSMUSG00000039167 & Eltd1 & EGF, latrophilin seven transmembrane domain containing 1\\
ENSMUSG00000004267 & Eno2 & enolase 2, gamma neuronal\\
ENSMUSG00000032314 & Etfa & electron transferring flavoprotein, alpha polypeptide\\
ENSMUSG00000010517 & Faf1 & Fas-associated factor 1\\
ENSMUSG00000025262 & Fam120c & family with sequence similarity 120, member C\\
ENSMUSG00000025153 & Fasn & fatty acid synthase\\
ENSMUSG00000070733 & Fryl & furry homolog-like (Drosophila)\\
ENSMUSG00000039801 & Gm5906 & RIKEN cDNA 2410089E03 gene\\
ENSMUSG00000031210 & Gpr165 & G protein-coupled receptor 165\\
ENSMUSG00000020176 & Grb10 & growth factor receptor bound protein 10\\
ENSMUSG00000030209 & Grin2b & glutamate receptor, ionotropic, NMDA2B (epsilon 2)\\
ENSMUSG00000031584 & Gsr & glutathione reductase\\
ENSMUSG00000006930 & Hap1 & huntingtin-associated protein 1\\
ENSMUSG00000029104 & Htt & huntingtin\\
ENSMUSG00000009828 & Ick & intestinal cell kinase\\
ENSMUSG00000023830 & Igf2r & insulin-like growth factor 2 receptor\\
ENSMUSG00000042599 & Jhdm1d & jumonji C domain-containing histone demethylase 1 homolog D (S. cerevisiae)\\
ENSMUSG00000024410 & K100 & RIKEN cDNA 3110002H16 gene\\
ENSMUSG00000016946 & Kctd5 & potassium channel tetramerisation domain containing 5\\
ENSMUSG00000063077 & Kif1b & kinesin family member 1B\\
ENSMUSG00000027550 & Lrrcc1 & leucine rich repeat and coiled-coil domain containing 1\\
ENSMUSG00000028649 & Macf1 & microtubule-actin crosslinking factor 1\\
ENSMUSG00000036278 & Macrod1 & MACRO domain containing 1\\
ENSMUSG00000008763 & Man1a2 & mannosidase, alpha, class 1A, member 2\\
ENSMUSG00000059474 & Mbtd1 & mbt domain containing 1\\
ENSMUSG00000020184 & Mdm2 & transformed mouse 3T3 cell double minute 2\\
ENSMUSG00000024294 & Mib1 & mindbomb homolog 1 (Drosophila)\\
ENSMUSG00000038056 & Mll3 & myeloid/lymphoid or mixed-lineage leukemia 3\\
ENSMUSG00000022889 & Mrpl39 & mitochondrial ribosomal protein L39\\
ENSMUSG00000033004 & Mycbp2 & MYC binding protein 2\\
ENSMUSG00000030739 & Myh14 & myosin, heavy polypeptide 14\\
ENSMUSG00000034593 & Myo5a & myosin VA\\
ENSMUSG00000027799 & Nbea & neurobeachin\\
ENSMUSG00000020716 & Nf1 & neurofibromatosis 1\\
ENSMUSG00000038495 & Otud7b & OTU domain containing 7B\\
ENSMUSG00000021140 & Pcnx & pecanex homolog (Drosophila)\\
ENSMUSG00000002265 & Peg3 & paternally expressed 3\\
ENSMUSG00000028085 & Pet112l & PET112-like (yeast)\\
ENSMUSG00000039943 & Plcb4 & phospholipase C, beta 4\\
ENSMUSG00000032827 & Ppp1r9a & protein phosphatase 1, regulatory (inhibitor) subunit 9A\\
ENSMUSG00000038976 & Ppp1r9b & protein phosphatase 1, regulatory subunit 9B\\
ENSMUSG00000003099 & Ppp5c & protein phosphatase 5, catalytic subunit\\
ENSMUSG00000039410 & Prdm16 & PR domain containing 16\\
ENSMUSG00000030465 & Psd3 & pleckstrin and Sec7 domain containing 3\\
ENSMUSG00000038764 & Ptpn3 & protein tyrosine phosphatase, non-receptor type 3\\
ENSMUSG00000053141 & Ptprt & protein tyrosine phosphatase, receptor type, T\\
ENSMUSG00000068748 & Ptprz1 & protein tyrosine phosphatase, receptor type Z, polypeptide 1\\
ENSMUSG00000037098 & Rab11fip3 & RAB11 family interacting protein 3 (class II)\\
ENSMUSG00000027652 & Ralgapb & Ral GTPase activating protein, beta subunit (non-catalytic)\\
ENSMUSG00000075376 & Rc3h2 & ring finger and CCCH-type zinc finger domains 2\\
ENSMUSG00000042453 & Reln & reelin\\
ENSMUSG00000050310 & Rictor & RPTOR independent companion of MTOR, complex 2\\
ENSMUSG00000020448 & Rnf185 & ring finger protein 185\\
ENSMUSG00000038685 & Rtel1 & regulator of telomere elongation helicase 1\\
ENSMUSG00000021313 & Ryr2 & ryanodine receptor 2, cardiac\\
ENSMUSG00000075318 & Scn2a1 & sodium channel, voltage-gated, type II, alpha 1\\
ENSMUSG00000028064 & Sema4a & sema domain, immunoglobulin domain (Ig), transmembrane domain (TM) and short cytoplasmic domain, (semaphorin) 4A\\
ENSMUSG00000005089 & Slc1a2 & solute carrier family 1 (glial high affinity glutamate transporter), member 2\\
ENSMUSG00000023032 & Slc4a8 & solute carrier family 4 (anion exchanger), member 8\\
ENSMUSG00000019769 & Syne1 & synaptic nuclear envelope 1\\
ENSMUSG00000062542 & Syt9 & synaptotagmin IX\\
ENSMUSG00000053580 & Tanc2 & tetratricopeptide repeat, ankyrin repeat and coiled-coil containing 2\\
ENSMUSG00000023923 & Tbc1d5 & TBC1 domain family, member 5\\
ENSMUSG00000039230 & Tbcd & tubulin-specific chaperone d\\
ENSMUSG00000032186 & Tmod2 & tropomodulin 2\\
ENSMUSG00000009470 & Tnpo1 & transportin 1\\
ENSMUSG00000019820 & Utrn & utrophin\\
ENSMUSG00000046230 & Vps13a & vacuolar protein sorting 13A (yeast)\\
ENSMUSG00000045962 & Wnk1 & WNK lysine deficient protein kinase 1\\
ENSMUSG00000047694 & Yipf6 & Yip1 domain family, member 6\\
ENSMUSG00000020812 &  & RIKEN cDNA 1810032O08\\
ENSMUSG00000053081 & & RIKEN cDNA 1700069B07\\
ENSMUSG00000072847 & & RIKEN cDNA A530017D24\\
\end{longtable}
\end{center}

\begin{table}[htbp]
\begin{center}
  \caption{23 genes with differential isoform usage (DU only p-value $<$ 0.01) between the two alleles of the haloperidol treated F1(129$\times$PWK) mouse,  but no differential isoform usage (DU only p-value $>$ 0.1) comparing the two alleles of the placebo treated F1(129$\times$PWK) mouse.}
  \vspace{1ex}
\begin{tabular}{l l p{8.5cm}}
\hline
\textbf{Ensembl ID} & \textbf{symbol} & \textbf{name} \\
\hline
ENSMUSG00000006638 & Abhd1 & abhydrolase domain containing 1\\
ENSMUSG00000005686 & Ampd3 & adenosine monophosphate deaminase 3\\ 
ENSMUSG00000004446 & Bid & BH3 interacting domain death agonist\\ 
ENSMUSG00000022617 & Chkb & choline kinase beta\\
ENSMUSG00000026816 & Gtf3c5 & general transcription factor IIIC, polypeptide 5\\ 
ENSMUSG00000031787 & Katnb1 & katanin p80 (WD40-containing) subunit B 1\\ 
ENSMUSG00000058740 & Kcnt1 & potassium channel, subfamily T, member 1\\ 
ENSMUSG00000039682 & Lap3 & leucine aminopeptidase 3\\ 
ENSMUSG00000026792 & Lrsam1 & RIKEN cDNA 4930555K19\\ 
ENSMUSG00000024085 & Man2a1 & mannosidase 2, alpha 1\\ 
ENSMUSG00000029822 & Osbpl3 & oxysterol binding protein-like 3\\
ENSMUSG00000021846 & Peli2 & pellino 2\\ 
ENSMUSG00000033628 & Pik3c3 & phosphoinositide-3-kinase, class 3\\ 
ENSMUSG00000005225 & Plekha8 & pleckstrin homology domain containing, family A (phosphoinositide binding specific) member 8\\ 
ENSMUSG00000026035 & Ppil3 & peptidylprolyl isomerase (cyclophilin)-like 3\\ 
ENSMUSG00000036202 & Rif1 & Rap1 interacting factor 1 homolog (yeast)\\ 
ENSMUSG00000001054 & Rmnd5b & required for meiotic nuclear division 5 homolog B (S. cerevisiae)\\
ENSMUSG00000052656 & Rnf103 & ring finger protein 103\\
ENSMUSG00000027273 & Snap25 & synaptosomal-associated protein 25\\ 
ENSMUSG00000043079 & Synpo & synaptopodin\\ 
ENSMUSG00000040389 & Wdr47 & WD repeat domain 47\\
ENSMUSG00000001017 & & RIKEN cDNA 2500003M10\\
ENSMUSG00000044600 & & RIKEN cDNA 9130011J15\\
\hline
\end{tabular}
\end{center}
\end{table}

\end{document}